\documentclass[11pt,oneside]{article}

\usepackage[breaklinks=true]{hyperref}
\usepackage{breakcites}

\usepackage{graphicx}
\usepackage[bottom=1.0in,margin=0.2in]{geometry}
\usepackage[]{hyperref}
\usepackage{booktabs}

\begin{document}
\newgeometry{bottom=1.5in}

\begin{flushright}
{\tt FERMILAB-PUB-20-649-ND} \\
Published in Harvard Data Science Review
\end{flushright}

\begin{center}

  {\Large{\bf{Reproducibility and Replication of Experimental Particle Physics Results}}}


  \vspace*{.2in}

  \begin{tabular}{cc}
    Thomas R. Junk$^{\dagger,*}$ and Louis Lyons$^{\ddag,**}$
   \\[0.25ex]
   {\small $^{\dagger}$ Fermi National Accelerator Laboratory, Batavia, IL USA} \\
   {\small $^{\ddag}$ Imperial College, London and Oxford University, UK}
  \end{tabular}

  \vspace*{0.2in}

  May 5, 2021

  \vspace*{0.4in}

\begin{abstract}
Recently, much attention has been focused on the replicability of scientific results, causing scientists, statisticians, and journal editors to examine closely their methodologies and publishing criteria.  Experimental particle physicists have been aware of the precursors of non-replicable research for many decades and have many safeguards to ensure that the published results are as reliable as possible.  The experiments require large investments of time and effort to design, construct, and operate.   Large collaborations produce and check the results, and many papers are signed by more than three thousand authors.  This paper gives an introduction to what experimental particle physics is and to some of the tools that are used to analyze the data.  It describes the procedures used to ensure that results can be computationally reproduced, both by collaborators and by non-collaborators.  It describes the status of publicly available data sets and analysis tools that aid in reproduction and recasting of experimental results.  It also describes methods particle physicists use to maximize the reliability of the results, which increases the probability that they can be replicated by other collaborations or even the same collaborations with more data and new personnel.   Examples of results that were later found to be false are given, both with failed replication attempts and one with alarmingly successful replications.  While some of the characteristics of particle physics experiments are unique, many of the procedures and techniques can be and are used in other fields.
\end{abstract}
\end{center}

\vspace*{0.15in}
\hspace{10pt}
  \small	
  \textbf{\textit{Keywords: }} {Reliability, Reproducibility, Replication, Particle Physics}
 
{\footnotesize This article is \copyright{} \the\year{} by author(s) as listed above. The article is licensed under a Creative Commons Attribution (CC BY 4.0) International license (https://creativecommons.org/licenses/by/4.0/legalcode), except where otherwise indicated with respect to particular material included in the article. The article should be attributed to the author(s) identified above.}

  \vspace*{0.4in}

\section*{Media Summary}

Scientists and statisticians alike in several fields are concerned by the low rate at which results have been confirmed when experiments are repeated.  Not every proposed solution makes sense, although there are many good ideas.  Particle physicists have long been aware of the precursors of non-replicable results.  Physicists in large collaborations who have worked long years on very expensive experiments do not wish to publish results that may later be found to be wrong, which would undermine the credibility of all results from that collaboration.   Thus, many internal tests of reproducibility of the results, as well as conservative methods such as blind analysis and stringent review, all with the purpose of catching mistakes and well-intentioned but flawed work, are common in particle physics.  Results are also published even if they disprove new theories; null results are not simply filed away.  Discoveries of new particles and interactions have a very high bar to meet in particle physics: $p$~values must be less than $2.87\times 10^{-7}$, not 0.05 as is common in some fields of study.  Particle physicists can easily point to past discovery claims that have had less significance and that have vanished when more data were collected or when other groups attempted confirmation.  Not every result is perfect or replicable in particle physics, but the quality is generally quite high.  New practitioners are always introduced to examples in which even the most careful analyzers have been able to fool themselves.  While some of the techniques and procedures used by particle physicists to ensure the reliability of their results are specific to the sub-field, many can be used regardless of the scientific specialty.


\section{Introduction}
\label{sec:introduction}

Experimental particle physics (EPP), also commonly known as high-energy physics, is a relatively mature field of research, with traditions spanning many decades.  Particle physics experiments take years to design, build and run, and they require large financial resources, equipment, and effort.  The ATLAS and CMS collaborations at the Large Hadron Collider (LHC) comprise more than 3000 physicists each.  Large collaboration size is not limited to collider physics -- neutrino experiments often have collaborations consisting of several hundreds to more than a thousand members.  Experimental particle physics has historically been on several cutting edges technologically, computationally, and sociologically.  This article gives a review of how issues of reproducibility and replication are addressed in the specific context of EPP.  Some of the techniques and procedures make use of distinctive features of EPP, while others are more broadly applicable.

Before addressing reproducibility and replication, however, we first give an introduction to some of the most important features of EPP.  Section~\ref{sec:WiPP} describes the target field of knowledge, elementary particle physics, and the tools used in the research: high-energy accelerators, particle detectors, data collection, processing, and reduction techniques.  Section~\ref{sec:data_analysis} describes common statistical inference tools used in EPP, including the estimation and inclusion of systematic uncertainties.  Section~\ref{sec:reproduction} defines what it means to reproduce an experimental result and describes some of the challenges and methods for addressing these in EPP. A summary of current efforts to preserve data and analysis workflows is given in this section.  Section~\ref{sec:replication} defines what it means to replicate an experimental result, with several EPP-specific nuances.  Section~\ref{sec:reliability} discusses the strategies used in EPP to maximize the reliability of the results.  Section~\ref{sec:specialfeatures} highlights aspects of EPP that differ from those of other fields of inquiry, motivating the choices of the procedures used.  Section~\ref{sec:examples_nonrep} gives examples of experimental results that have been not been replicated, and  Section~\ref{sec:example_toomuchrep} gives an example of a result that was replicated but was later found to be incorrect nonetheless.  Section~\ref{sec:conclusion} summarizes our conclusions.  Appendix~\ref{appendix:glossary} contains a glossary of particle physics terminology used in this paper.

\section{What is Particle Physics?}
\label{sec:WiPP}

Particle physics involves the study of matter and energy at its smallest and most fundamental level. A short introduction is given below to the current understanding of known elementary particles and the forces with which they interact, as well as a brief description of the equipment used to perform particle physics experiments.

\subsection{Particles and Forces}
\label{sec:particlesandforces}

The question of what are the smallest building blocks out of which everything is made has
a long history. For the ancient Greeks, the elements were Air, Fire, Earth and Water.

Today's elementary particles are quarks and leptons, all with spin 1/2; each has its own antiparticle.
The leptons comprise the electron and its two heavier versions, the muon ($\mu$)
and the heavy lepton ($\tau$);  for each of these, there is a corresponding uncharged neutrino. 
The proton, the neutron and other half-integer spin particles (baryons) are 
composed of combinations of three quarks; for example the neutron consists of an up quark and two down quarks. Mesons of integer spin 
(bosons) consist of a quark and an antiquark.  

The baryons and mesons, the particles made of quarks, are collectively known as hadrons.   Quarks are confined within hadrons, and in contrast to leptons do not seem to lead an independent existence as free particles. In collisions between particles, as any struck quarks try to escape from the hadrons, they are converted into jets of pions, kaons and protons which leave visible tracks in the detectors (see Figure~\ref{fig:cmshzzevd}). 

In addition, the fundamental forces are each transmitted by their own carrier. These are
\begin{itemize}
\item{Gravitation, transmitted via gravitons.   Gravitons, the quanta of the gravitational field, have not yet been observed and it may be very difficult to do so (\cite{Rothman:2006fp}).  Gravity is a ``long-range'' force.  ``Short-range'' forces, on the other hand, are confined to distances on the scale of the size of an atomic nucleus.  Gravitational waves produced in the coalescence of a pair of black
holes were observed in 2016 (\cite{Abbott:2016nmj}).}
\item{Electromagnetism. This long-range force is transmitted by photons.}
\item{The weak nuclear force. The intermediate vector bosons $W^+$, $W^-$ and $Z^0$ transmit this short-range force.  The weak nuclear force has a unified description with the electromagnetic force -- the resulting description is called the ``electroweak'' model.  The $W^\pm$ and $Z^0$ bosons were discovered in 1983 by the UA1 and UA2 collaborations at the CERN SPS collider (\cite{Arnison:1983rp}; \cite{Banner:1983jy}; \cite{Arnison:1983mk}; \cite{Bagnaia:1983zx}). Thomson (\cite{Thomson:2013zua}) provides a modern review of this interaction.}
\item{The strong nuclear force. This is another short range-force, transmitted by gluons. They are
responsible for binding quarks in hadrons, are  produced in
collisions involving quarks, and are detected by the jets of particles they produce.  The gluon was discovered in 1979 by the TASSO, Mark-J, PLUTO and JADE collaborations at the PETRA $e^+e^-$ collider at the Deutsches Elektronen-Synchrotron Laboratory   (\cite{Brandelik:1979bd}; \cite{Barber:1979yr}; \cite{Berger:1979cj}; \cite{Bartel:1979ut}). The review by Huston, Rabbertz, and Zanderighi (\cite{Zyla:2020zbs}) provides details of the current state of knowledge of this force.}
\end{itemize}  
Finally, there is the Higgs boson, with a mass of approximately 125 GeV. \footnote{A GeV is a unit of energy or mass, and is $1\times 10^9$ electron volts.  For comparison, the lightest neutrino's mass is less than 1.1~eV (\cite{Aker:2019uuj}), the proton's is about 1 GeV, and a gold atom's is about 180 GeV.} The Higgs field is responsible for enabling fundamental particles to have mass.  Without the Higgs field or something that takes its place, the symmetries of the electroweak model predict elementary particles to be massless.  The Higgs boson was discovered in 2012 by the ATLAS  and CMS collaborations at the LHC (\cite{Aad:2012tfa}; \cite{Chatrchyan:2012ufa}).

In the standard model (SM) of particle physics, quarks and leptons are arranged in three generations of increasing mass - see
Table~\ref{Table:generations}. It is not understood why there are three
generations.  In each generation there is a quark with charge $+2e/3$ (the ``up-type'' quark), one with charge $-e/3$ (the ``down-type'' quark), a lepton with charge $-e$, and a corresponding neutrino.  The neutrino states mix so that the states with definite mass are not the ones with definite flavor (that is, $\nu_e$, $\nu_\mu$, or $\nu_\tau$) but rather they are linear combinations of the three flavor states.  The three neutrino masses are all very small; two of them are close in mass, but it is not known whether the third is heavier
(``normal mass ordering'') or lighter (``inverted ordering'') (\cite{deSalas:2018bym}; \cite{Esteban:2018azc}).  Neutrinos are produced and detected not in their mass states but rather in their flavor states.  The mass states however propagate unchanged over time according to the rules of quantum mechanics.  Neutrinos produced in definite flavor states therefore ``oscillate'' in flavor among the $\nu_e$, $\nu_\mu$, and $\nu_\tau$ states as they travel.  Gonzalez-Garcia and Yokoyama give an extensive review of the phenomenon~(\cite{Zyla:2020zbs}).

The SM also specifies the way the various particles and force carriers interact
with each other via the fundamental interactions in the bottom three rows of Table~\ref{Table:Forces}. Gravitation has not yet been unified with the other three forces. It has not been included in the SM, and it has a negligible effect for most particle physics measurements. 

Models describing particles and interactions that are not included in the SM are termed ``beyond the standard model'' (BSM).  One of the primary reasons to run particle physics experiments is to test these models.  Both exclusions and discoveries advance the state of knowledge.  One class of BSM models that is used to provide examples in this paper is supersymmetric models.  These models predict the existence of a supersymmetric partner for every particle in the SM.  
SM particles with half-odd integer spin, called fermions, have bosonic
supersymmetric partners with integer spin and names that
are constructed from the SM particles’ names, prepended with an ``s,''
and SM bosons have fermionic supersymmetric partners with names that
end in ``-ino.'' Examples are the gravitino $\widetilde G$, which is the superpartner of the graviton, and the selectron, the superpartner of the electron.  Supersymmetric models have additional Higgs bosons, including charged Higgs bosons $H^\pm$.

\begin{table}
\begin{center}
\caption{Fundamental particles: quarks and leptons. The electric charge is in units of the charge on a proton ($e$). Each has its own antiparticle.} \label{Table:generations}
\vspace{0.3cm}
\begin{tabular} {| c | c |  c | c  | c   |}
\hline
& Quark & Quark& Lepton    &Lepton \\ \hline
Generation 1 & Up $u$ &  Down $d$  &  Electron $e$ & Neutrino $\nu_e$ \\ 
Generation 2 & Charm $c$ & Strange $s$  &  Muon $\mu$ & Neutrino $\nu_{\mu}$ \\ 
Generation 3 & Top $t$ & Bottom $b$  &  Tau $\tau$ & Neutrino $\nu_{\tau}$ \\   \hline 
Electric charge  & $+2e/3$   &  $-e/3$  &    $-e$     &    0  \\ \hline
\end{tabular}
\end{center}
\end{table}

\begin{table}
\begin{center}
\caption{Fundamental forces}
\label{Table:Forces}
\vspace{0.3cm}
\begin{tabular} {  |   c   | c | c  |  c | c  |}
\hline
Force &  Potential  &  Force carrier &  Responsible for  &   Particles affected  \\  \hline
Gravitation  \rule{0pt}{5mm}     &  $1/r$ &   Graviton    &  Earth going     & Everything    \\   
                &        &            &  round Sun, etc.           &   \\
Electro-  \rule{0pt}{6mm}       &    $1/r$    &   Photon    &    Coulomb repulsion,  &   Charged particles\\ 
magnetism        &             &             &          photon emission   &        \\
Weak \rule{0pt}{6mm}           &    Short range     & $W^\pm$ and $Z^0$ & Energy generation    &  Quarks, Leptons\\    
                  &                   & bosons   &    in Sun, $\beta$-decay &            \\
Strong \rule{0pt}{6mm}    & Short range   &     Gluons   & Nuclear binding,    &    Hadrons   \\
  \rule[-3mm]{0pt}{2mm}        &                   &              & quark-quark scattering                   &            \\  \hline
\end{tabular}
\end{center}
\end{table}

The remainder of this section describes how the data are produced, recorded, and processed for use in physics analyses.  Particle accelerators and detectors are the apparatus used to create and record the interactions to be studied.  The choices made in their design affect the types of studies that can be carried out and their ultimate sensitivities. 

\subsection{Accelerators}
Experiments can be divided into those performed at accelerators, and those carried out elsewhere. The
accelerator ones either use a beam hitting a stationary target; or have antiparallel beams colliding with each other.
An example of the former is a neutrino beam, with the detector hundreds of kilometers away (\cite{Anderson:1998zza}; \cite{ABE2011106}; \cite{Abi:2020wmh}).  
Colliding beams provide an easier way of achieving higher center-of-mass energies, but they have more stringent beam 
requirements. The highest center-of-mass energy of 13~TeV has been obtained at CERN's LHC with collisions between protons in a 27~km circumference ring of superconducting magnets some 100 - 150 m below the surface (\cite{Evans:2008zzb}). A typical 
analysis uses data collected over a running period of between several weeks and several years.  

The size of a data set collected by an experiment and used for an analysis is usually expressed in terms of how many events are expected on average in a data set of that size.  A standard unit of cross section is a ``barn,'' which is roughly the cross sectional area of a uranium nucleus.  Rarer processes have cross sections more conveniently expressed in millibarns, microbarns, nanobarns, picobarns, and femtobarns.  The term used for the data set size is ``integrated luminosity,'' and it is typically reported in inverse femtobarns (fb$^{-1}$) at the LHC.  A process with a cross section of 1~fb is expected to happen once, on average in a data set size of 1~fb$^{-1}$.

There are various forms of non-accelerator experiments.  Nuclear reactors are sources of particles, such as neutrons and neutrinos, that are the subjects of many experiments.  Cosmic rays are another source of high-energy particles for performing astrophysics and particle-physics experiments.  Measurements of solar and atmospheric neutrinos and many searches for dark matter, proton decay and other exotic phenomena likewise do not require the use of accelerators. 

\begin{figure}[ht]
\centering
\includegraphics[width=\textwidth]{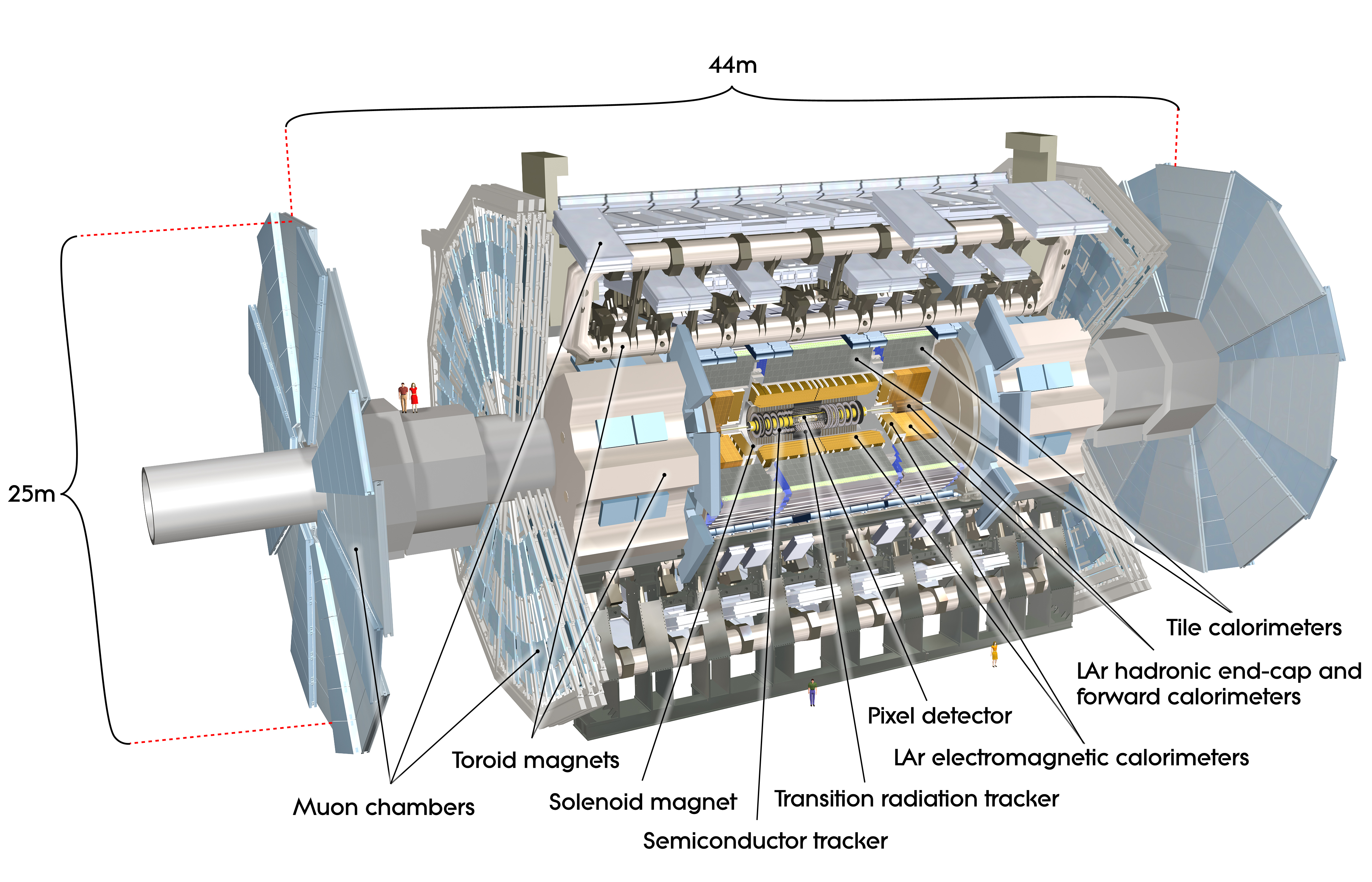} \\
\caption{Cutaway view of the ATLAS particle detector at the LHC.  Four human figures are drawn to emphasize the scale.  The total weight of the detector is approximately 7000 tons.  It has approximately 100~million independent readout channels and 3000~km of cables.  Appendix~\ref{appendix:glossary} provides short descriptions of the named detector components.  Credit: CERN.}
\label{fig:atlas}
\end{figure}

\subsection{Detectors}
There are four large, general-purpose detectors at the LHC: ATLAS (\cite{Aad:2008zzm}), CMS (\cite{Chatrchyan:2008aa}),  ALICE (\cite{Aamodt:2008zz}) and LHCb (\cite{Alves:2008zz}).  
Each of these four detectors consists
of sub-detectors arranged around the accelerator's beam pipe. 
Figure~\ref{fig:atlas} shows a cutaway view of the ATLAS detector.
Its sub-detectors have different functions:
\begin{itemize}
\item{Pixel vertex detector. This is a high spatial resolution detector placed as close as possible to the interaction region. It is 
useful for finding charged particles that come from the decays of heavier particles that have traveled millimeters before decaying.}
\item{Tracker. This detects charged particles and measures their momenta.}
\item{Electromagnetic calorimeter. Electrons and photons are identified by the showers of $e^+e^-$ pairs and photons they produce in the calorimeter's
material, which contains heavy elements such as lead or tungsten.}
\item{Hadron calorimeter.  Located outside the electromagnetic calorimeter, this sub-detector measures the energy left by pions, kaons, neutrons, and other neutral and charged hadrons.  Hadron calorimeters are usually made of layers of steel with particle-detecting layers sandwiched in between.}
\item{Muon detector. These are placed at the outside of the whole detector, so that nearly all charged particles apart from muons are absorbed before reaching these detectors.}
\end{itemize}
These are in a magnetic field of several Tesla, so that the momenta of charged particles can be measured. 
Because these are general-purpose detectors, many different physics analyses are possible. Usually these will be performed on different subsets of the accumulated data.

Experiments at lower energy accelerators tend to have more individually designed, smaller detectors for their specific analysis. 

\subsection{Trigger}
The intensities of the  proton beams at the LHC are such that the rate of collisions at the center of ATLAS and CMS is of order $10^9$ Hz. The data acquisition system can  record up to 1000 interactions (``events'') per second.  An online trigger system is used to select interesting events to be stored for further analysis; this consists of several algorithms in parallel, to cater for the variety  of the subsequent physics analyses. Events not recorded are lost.
Studies are performed to evaluate and correct for the trigger efficiency for recording wanted events for each analysis.

\subsection{Reconstruction} 
\begin{figure}[ht]
    \centering
    \includegraphics[scale=0.35]{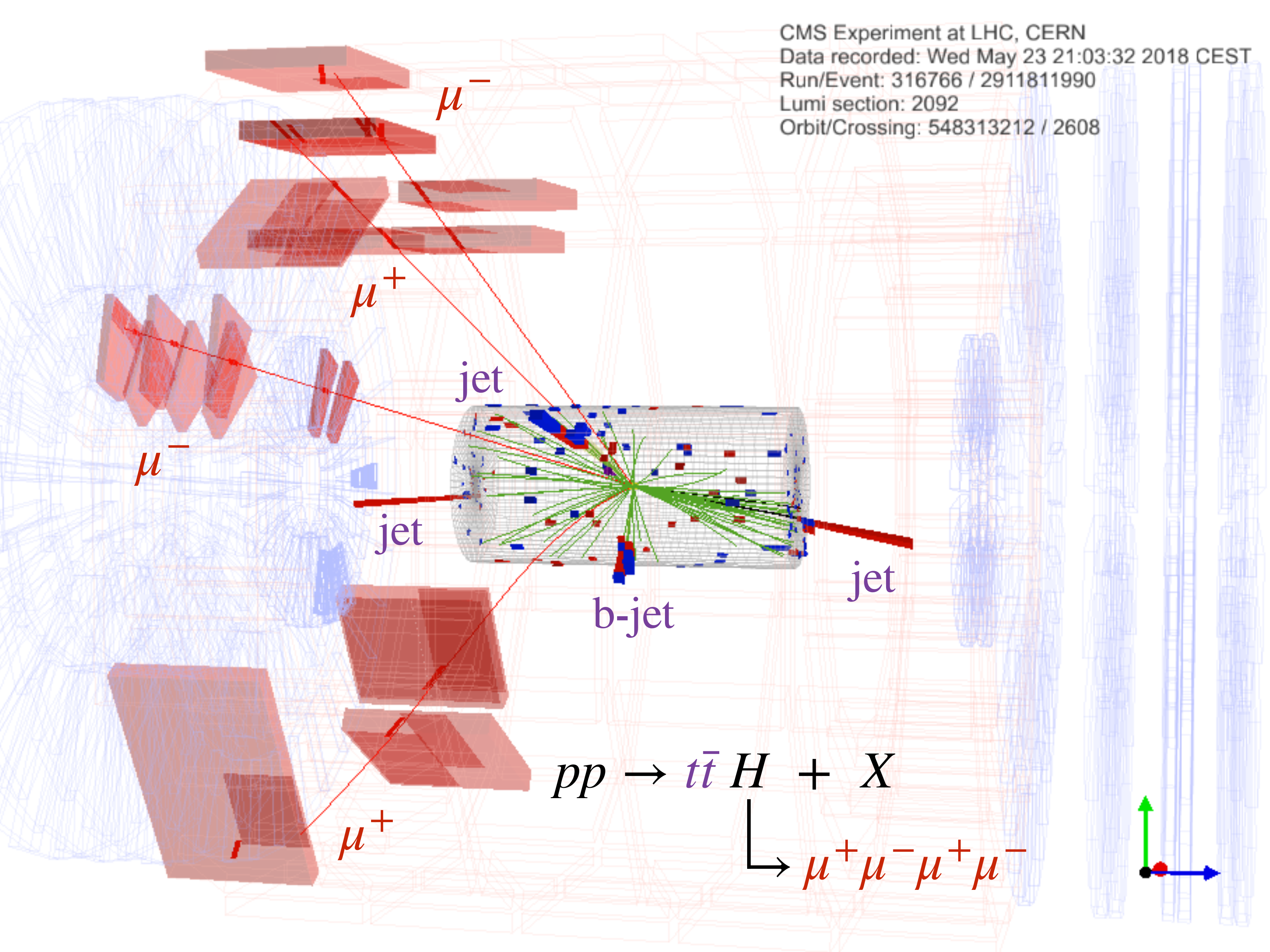} \\
    \caption{An event display from the CMS collaboration, showing a $pp\rightarrow t{\bar{t}}H+X$ candidate interaction. In a collision of two protons ($pp$), a top-antitop-quark pair ($t{\bar{t}}$) is produced, along with a Higgs boson ($H$) and other particles ($X$). The Higgs boson decays to two $Z^0$ bosons which themselves each decay to $\mu^+\mu^-$.  The top quarks decay to multiple jets of hadrons.  These decays are very rapid -- the Higgs boson, the top quarks, and the $Z^0$ bosons travel distances that are too small to be measured before they decay, but their presence can be inferred from their visible decay products.  Reconstructed tracks are shown as curves that originate in the center of the detector, and calorimeter clusters and muon detector responses are shown with shaded blocks farther out.}
    \label{fig:cmshzzevd}
\end{figure}

The information from the detectors consists of a series of digitized electronic signals which measure energy deposits (``hits'') in small regions of known location in the detector. The job of the reconstruction programs is to link together appropriate hits and turn these into a series of three-dimensional tracks corresponding to the trajectories  particles took after they were produced in the event\footnote{In a region where the magnetic field is constant, the trajectories of charged particles are approximately helical. Neutral particles travel in straight lines.}.  Figure~\ref{fig:cmshzzevd} shows the reconstructed tracks and calorimeter clusters for a single event collected by the CMS detector. This event passes the candidate selection requirements for the production in proton-proton collisions of a Higgs boson and two top quarks (\cite{CMS:2020dkv}).       
   
    \section{Data Analysis in Particle Physics}
\label{sec:data_analysis}
In a large experiment such as a collider or a modern neutrino experiment, the reconstructed data are shared among all collaborators.  Physicists group together to perform data analyses which produce publishable results.  Physics analyses involve selecting events, comparing the observed distributions of events with those predicted by models, conducting statistical inference, and estimating and including systematic uncertainties in the final results.  The issues discussed below relate to EPP, but the discussion is broadly applicable to many other fields of science.
 
 \subsection{Event Selection}
In general, each physics analysis will use a small subset of the accumulated data to reduce background from unwanted processes while maintaining a high efficiency for retaining the signal. The initial stage of this process involves relatively simple selection criteria devised by physicists (for example, the event should contain a muon and an electron of opposite electrical charge). This is usually followed by machine learning (ML) techniques, such as neural networks or boosted decision trees. Recently, deep learning methods have been employed.  Carleo and collaborators provide a recent review of ML techniques in several sub-fields of physical sciences, including EPP (\cite{Carleo:2019ptp}). When ML tools are used, the choice of training samples is important.  

\subsection{Statistical Interpretation} 
\label{sec:statinterp}

There are two large classes of analyses in EPP, parameter determination and hypothesis testing.

{\it Parameter Determination.}  This involves the determination of one or more parameters (e.g. the mass of the Higgs boson) and the associated uncertainties or parameter ranges at some specified confidence level. It requires the use of some parameter determination technique, such as those  listed below, each of which exists in several variants.

\begin{itemize}
    \item{Chi-squared: This can be the Pearson or the Neyman versions, or can use a log likelihood ratio formulation. } 
    \item{Likelihood: The data can consist of individual events, or they can be accumulated in the bins of a histogram. The usual form of the likelihood uses just the shape of the expected distribution of data for different values of the parameter(s), with the overall normalisation kept fixed.  The extended form allows the overall number of events to fluctuate. For histogrammed data, these correspond to the numbers of events in the bins being multinomially distributed, or to have independent Poisson distributions, respectively.}
    \item{Bayesian posterior: Although Bayesian methods for hypothesis testing are not often used in experimental particle physics, they are more often used for parameter determination. The choices here are the functional forms used for the Bayesian priors, and the way the credible interval is extracted from the posterior.}
    \item{Neyman construction: This guarantees confidence coverage\footnote{This is a property of the statistical procedure used to determine a range for a physical parameter. If the experiment were to be repeated many times, these ranges would vary because of statistical fluctuations. Coverage specifies the probability of these ranges containing the true value of the parameter.}  for the determined parameter(s). The resulting confidence intervals can be chosen to be one-sided    
    upper limits (UL) or lower limits (LL); two-sided central intervals; or Feldman-Cousins (\cite{Feldman:1997qc})}.
    For constructing the confidence belt at a given value of the parameter, Feldman and Cousins use a likelihood-ratio ordering for the data, as implied by the theory of likelihood ratio tests (see Sect 23.1 of \cite{kendallstuartord}). 
\end{itemize}

As well as determining the actual confidence interval from the data, it is also important to calculate the distribution of intervals that are expected.  Sensitivities are often quoted using the median interval bounds from a large number of simulated repetitions of the experiment assuming the data are distributed according to the relevant model; or from the ``Asimov'' data set, where the single set of invented ``data'' is exactly as predicted by the relevant model (\cite{Cowan:2010js}).

\begin{figure}[ht]
\centering
\includegraphics[scale=0.35]{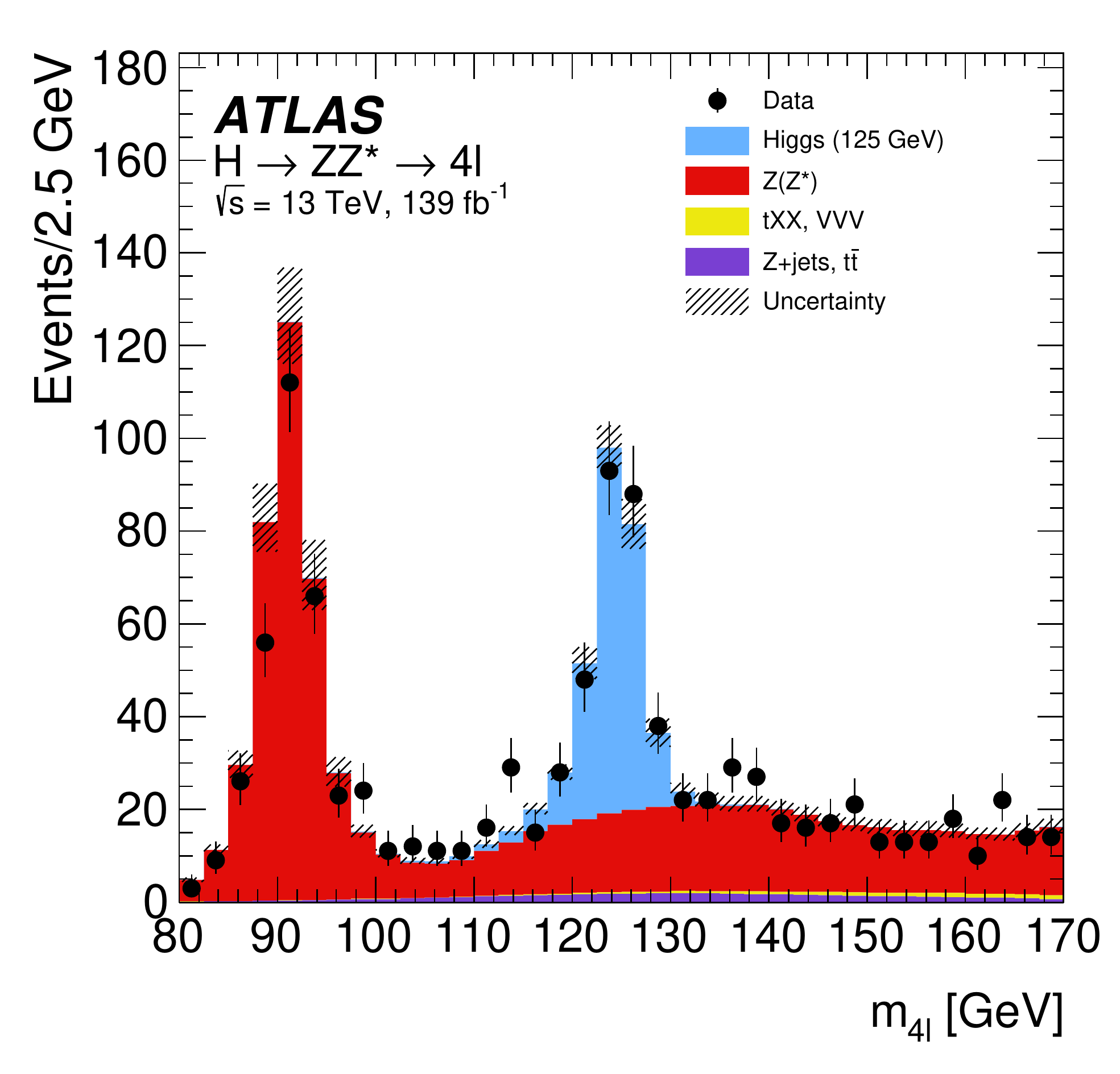} \\
\caption{A histogram showing event counts in bins of the reconstructed mass $m_{4\ell}$ for interactions selected in the $H\rightarrow ZZ^*\rightarrow 4\ell$ decay mode in the ATLAS detector (\cite{ATLAS:2020wny}). The points with error bars show the numbers of observed interactions in each bin, and the shaded, stacked histograms show the model predictions. The red peak on the left corresponds to the well-known $Z^0$ boson (an example of a ``standard candle''), while the blue peak in the middle shows the prediction for the Higgs boson.  A $Z^*$ is a $Z^0$ boson with an invariant mass that differs from the most probable mass of the $Z^0$ boson.  The $tXX$ prediction includes $t{\bar{t}}Z$, $t{\bar{t}}W$, $tWZ$, and other rare processes containing top quarks.  The $VVV$ prediction includes $WZZ$, $WWZ$, and  other tri-boson contributions. }
\label{fig:atlashzz4l}
\end{figure}

{\it Hypothesis Testing.}  This is where we attempt to see if the data favors some version of new physics (hypothesis $H_1$) as compared with a null hypothesis $H_0$. Thus if we had a mass spectrum and were looking for a peak at some location in it (see Fig~\ref{fig:atlashzz4l}),
 our hypotheses could be:
\begin{itemize}
    \item{$H_0$ = only well known particles are produced.}
    \item {$H_1$ = also the production of Higgs bosons, decaying via a pair of $Z^0$ bosons to four charged leptons.}
\end{itemize}  
Alternatively, an example from neutrino physics would be
\begin{itemize}
    \item {$H_0$ = ``normal'' ordering of the three neutrino masses, or}
    \item{$H_1$ = ``inverted'' ordering.}
\end{itemize}

Here some form of hypothesis testing is used. This usually requires the choice of a data statistic, which may well be a likelihood ratio under the two hypotheses. Then the choice of hypothesis favored by the data may involve comparing the actual value of the data statistic, with the expected distributions under the two hypotheses; these may be obtained by Monte Carlo simulation. Another possibility is to use the expected asymptotic distributions (\cite{Cowan:2010js}), though care must be taken to use the asymptotic formulas  only within their domains of applicability.

\begin{figure}[ht]
    \centering
    \includegraphics[scale=0.75]{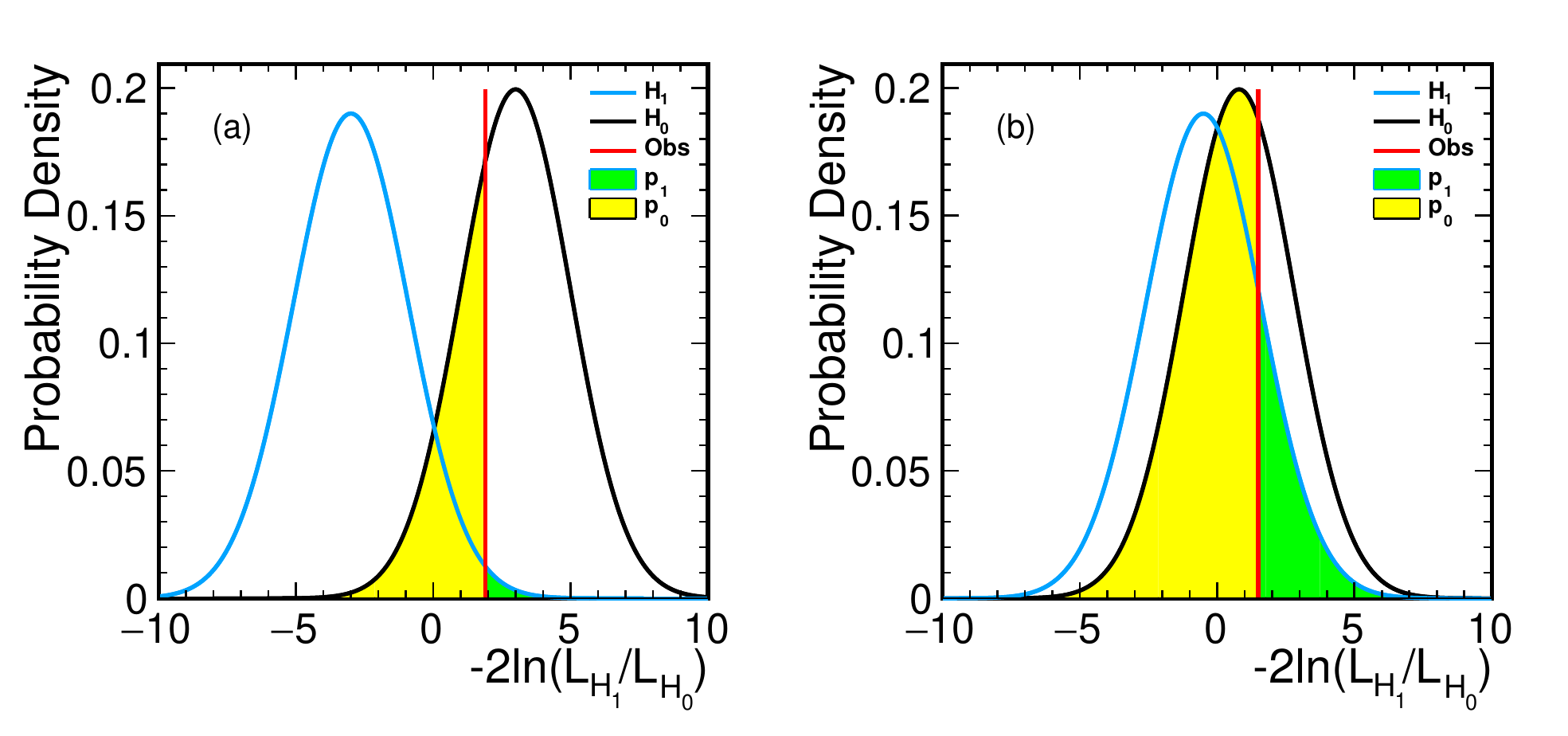} \\
    \caption{Example distributions of the logarithm of the likelihood ratio test statistic assuming $H_0$ (black) and $H_1$ (blue).  An example observed outcome of the experiment is indicated by the red line.  The $p$~value $p_0$ is the yellow-shaded area under the tail of the $H_0$ distribution to the left of the observed value, and $p_1$ is the green-shaded area under the $H_1$ distribution to the right of the observed value.  Panel (a) shows a case in which the experiment is expected to distinguish between $H_0$ and $H_1$ most of the time, and panel (b) shows the distributions for an experiment that is not as sensitive.}
    \label{fig:p0p1}
\end{figure}

Possible outcomes of these comparisons are:
\begin{itemize}
\item{Data are consistent with $H_1$ but not with $H_0$. For the first example shown, this would constitute evidence for a discovery claim.}
\item{Data are consistent with $H_0$, but not with $H_1$. This would result in the exclusion of the model of new physics, at least for some values of the parameters of that model.  Figure~\ref{fig:p0p1}a shows an example of this situation.} 
\item{Data are consistent with both $H_0$ and $H_1$.  The experiment is not sensitive enough to distinguish between the two models.  Figure~\ref{fig:p0p1}b is an example of this situation.} 
\item{Data are inconsistent with both $H_0$ and $H_1$. This may indicate that some other model is required.}
\end{itemize}

Particle physics analyses often result in the second situation above.  Publication of such null results is useful in that it serves to exclude the tested model of new physics at some confidence level, at least for some range of its parameter space. For example, selectrons are excluded at the $95\%$ confidence level for masses below 500~GeV, although this is somewhat model dependent (\cite{Aaboud:2018jiw}). That means that if the mass of the selectron were lighter, it would have produced a clear signal in the data, but this was not seen, so selectron masses below 500~GeV are ruled out.
An additional reason for publishing null results is that it avoids the publication bias of  accepting for publication only positive results.

\subsection{Systematic Uncertainties}  
\label{sec:systematic_uncertainties}
    
We define the ``error'' in a measurement to be the difference between the estimate of a model parameter produced by the analysis and the unknown true value.  ``Uncertainties'' are numerical estimates of the possible values of the errors, and are typically reported as one-standard-deviation intervals centered on the best estimate of a measured quantity.  Asymmetric confidence intervals are also used when appropriate (\cite{Barlow:2003sg}; \cite{Barlow:2004wg}).  Particle physicists often use the word ``error'' when they mean ``uncertainty.''
    
Physics analyses are affected by statistical uncertainties and by systematic ones. The former arise either from the limited precision of the apparatus and/or observer in making measurements, or from the random fluctuations (usually Poissonian) in counted events. They can be detected by the fact that, if the experiment is repeated several times, the measured physical quantity will vary.
    
Systematic effects, however, can cause the result to be shifted from its true value, but in a way that does not necessarily change from measurement to measurement. Measurements nearly always have some bias, and the question is by how much are they biased.  Results are corrected for known biases, and the uncertainties in these corrections contribute to the total uncertainties.  Systematic effects are not easy to detect, and in general much more effort is needed to evaluate the corresponding uncertainties.
    
The easiest sources of systematic uncertainty to estimate arise from propagating uncertainties from subsidiary measurements that constrain nuisance parameters.  These subsidiary measurements themselves have statistical and systematic uncertainties associated with them, and it is important to account for sources of systematic uncertainty that are common to the subsidiary and the main experiment, as well as between subsidiary measurements if there are more than one, with shared nuisance parameters.  Typical sources of systematic uncertainty that are evaluated in this manner arise from modeling the detector response.
    
For the case of systematic uncertainties that are constrained by subsidiary measurements, there is usually an extrapolation uncertainty which needs to be included, to account for errors in the assumptions that relate the result in the subsidiary measurement to the desired input to the measurement of interest.  These extrapolations often involve theoretical inputs, such as estimations of sample composition or distributions of properties of events that may differ between those selected in the subsidiary measurement and the main measurement.
    
The uncertainties on theoretical predictions are evaluated in one or more of the following ways.  A theoretical prediction may depend on other subsidiary measurements, in which case uncertainties are propagated as before.  Predictions by different teams of theorists who make different approximations or who take different calculational approaches are compared and the differences assigned to be the systematic uncertainty.  Sometimes this latter approach is criticized as the true value of what is being predicted may not be contained in the range of available predictions; the predictions may suffer from common errors.  The accuracy of some theoretical predictions is limited by the number of terms taken in difficult-to-calculate Taylor expansions.  Approximations are made to account for the possible values the missing higher-order terms may have.  Guidance is taken from similar calculations computed with available higher-order terms compared with truncated series.  Sometimes, for very conservative estimates of systematic uncertainties used for preliminary results, models which clearly are missing known physical processes are used as guesses to limit how big an uncertainty can be.
    
Propagating systematic uncertainties through an analysis often requires repeating Monte Carlo simulations with varied values of the nuisance parameters, be they associated with the theory or the modeling of the apparatus.  Monte Carlo samples must be large enough so that the statistical uncertainties in their predictions do not overwhelm the systematic effects under study (\cite{Barlow:2002yb}).
    
Inclusion of systematic uncertainties in the calculation of the significance of effects or the sizes of confidence intervals is necessary for the results in EPP to be meaningful.  Usually systematic errors are dealt with in  a likelihood function  by assigning them nuisance parameters, with constraint terms corresponding to the uncertainties on their values. Common ways of including their effects in final results such as $p$~values and confidence intervals are to profile the likelihood function with respect to them, or to marginalize the posterior probability distribution in a Bayesian approach.  The procedure recommended by the LHC Higgs Combination Group is to fix the values of the nuisance parameters when generating simulated experimental outcomes to those values that best fit the data collected by the experiment.  When performing the maximum-likelihood fits in the calculation of the profile likelihood ratio test statistic, the constraints on the nuisance parameter values are modeled as the outcomes of (possibly fictitious) auxiliary experiments, whose values fluctuate from one simulated experiment to the next (\cite{ATLAS:2011tau}; \cite{Cowan:2010js}).

The abstract and the conclusions  of an EPP paper will typically quote a result as $\mu \pm \sigma_1 \pm \sigma_2$, where $\mu$  is the measured quantity, $\sigma_1$ is the statistical uncertainty  and $\sigma_2$ is the systematic one.  No correlation is implied between the two sources of uncertainty, and they are usually considered to be independent.  The separation of the two uncertainties conveys to the reader both the claimed accuracy of the result and the claimed precision.  Because the statistical uncertainty is expected to become smaller as more data are collected while systematic uncertainties have constant components, the reader may judge whether it makes sense to run the experiment longer in order to improve the total uncertainty, or if the result is ``systematics limited.''  
 
 In the bulk of a paper, the numerical effects on the total systematic uncertainty are quoted separately for each source.
 The impact of including a result in a combination with other measurements of the same quantity also depends on the relative amounts of statistical and systematic uncertainty, the contributions from each source of uncertainty, and the correlations between the sources of uncertainty among the measurements being combined.   The LEP Electroweak Working Group's joint fits provide excellent examples of this in action (\cite{ALEPH:2005ab}).  Subsequent reinterpretation of results, especially if they are not replicated, is aided by a complete specification of the systematic uncertainties.  Meyer, Betancourt, Gran and Hill provided a recent example of a reinterpretation of bubble-chamber results of neutrino interactions on deuterium nuclei collected by independent experiments in the 1980s in which model assumptions and systematic uncertainties have been re-evaluated (\cite{Meyer:2016oeg}).
 
Experimentalists interested in improving the accuracy of the estimation of a parameter of interest but who do not have access to the equipment to replicate the experiment may choose instead to run an experiment that constrains one or more nuisance parameters.  The degree to which this additional experiment is worthwhile depends on the contributions of the systematic uncertainty to the original measurement.  
 
Usually an analysis will have optimizable parameters, such as data selection requirements, the adjustment of which creates a tradeoff between the sizes of the statistical and systematic uncertainties.  Optimizing these parameters for a standalone measurement may not produce the same optimization as for the case in which the measurement is combined with others or if the experiment is run longer.  Many analyses divide the experimental data into finely-binned subsamples, each of which produces its own measurement of the desired parameter, but the combination of all of them provides the best estimate.  Subsamples that are the least affected by systematic uncertainties often encompass small subsets of the data, as they select well-understood decay modes with small backgrounds, for example, but at a cost of smaller expected signals.  As the accelerator runs longer, these subsamples dominate the parameter estimate.  Furthermore, in situ constraints of nuisance parameters using control samples (see Section~\ref{sec:controlsamples}) reduces uncertainties that are nominally classified as systematic but which really have a statistical component.
 
For these reasons, the quoted systematic uncertainties on a measured parameter usually shrink as more data are collected, though often more slowly than the statistical uncertainties.  Further effort on the part of the experimental collaboration, other experiments, and the theory community also brings down systematic uncertainties over time.  Using older results, published when the understanding of systematic effects was less complete than it is at the time the results are used as input to future work, involves steps to correct and adjust for the new knowledge.  Detailed breakdowns of systematic uncertainties by their source and full explanations of how they were estimated assists in this process.
 
 The most problematic sources of systematic error are those that are neglected or incorrectly dismissed.  Sometimes they are simply not appreciated at the time the experiment is designed (``unknown unknowns'').  Experiments are often upgraded or run in different configurations in order to address these.  If systematic uncertainties remain difficult to control, a solution can be to perform a measurement of a quantity that can be more unambiguously defined, such as the ratio of two production rates.  Many systematic errors, both known and unknown, are expected to cancel in such ratios.  Some analyses claim to be ``model-independent,'' though often this phrase just means less dependent on models than similar measurements.
    
More details on the subject of systematic uncertainties can be found in Section~\ref{sec:review} and in reviews by Heinrich and Lyons (\cite{Heinrich:2007zza}) and Barlow (\cite{Barlow:2002yb}).  Systematic uncertainties are common in other fields of study as well.  For example, Meng (\cite{meng2018}) studies in detail the effects of systematic biases in population surveys.
      
 \subsection{{\it P}~Values}
 \label{sec:pvalues}
 
 
 Recently, $p$~values have been under attack (\cite{doi:10.1080/00031305.2018.1527253}), with some journals actually banning their use (\cite{journalbanspvalues}). The reasons seem to be:
 \begin{itemize}
\item{There are many results that claim to observe effects, based on having a $p$~value less than 0.05, which are subsequently not replicated.}
\item{People confuse the $p$~value with the probability of the null hypothesis being true. For example, it is often remarked that scientists analyzing data don't know what they are doing; when using $p_0 <5 \%$ as a discovery criterion, they are wrong with far more than $5\%$ of these discovery claims. This criticism is logically incorrect, and assumes that Prob(A|B) = Prob(B|A), which is generally not true\footnote{LL's granddaughter provided the example that the probability of eating toast for breakfast, given the fact that a person is a murderer, is about $50\%$, while the probability of being a murderer, given the fact that toast is part of the breakfast menu, is (thankfully) considerably smaller.}. The $p$-value is about Prob($p_0 < 5\%$ | given $H_0$ is true), while the  faulty criticism involves Prob($H_0$ is true | given $p_0  <5\%$). As an example, consider checking special relativity in LHC events, and using a (ridiculously low)  cut-off of $5\%$ for the p-value for $H_0$ ( =  special relativity is true).  Assuming special relativity is true, about $5\%$ of the tests should fail this criterion simply because of fluctuations, and all of these tests would be ``wrong.'' But there is nothing incorrect or paradoxical about this. Possible evidence against special relativity would require the fraction of $H_0$ tests failing the $p_0$ criterion to be significantly larger than $5\%$.}
 \end{itemize}
 The first point can be mitigated by having a lower cutoff on the $p$~value criterion. The second argument can be addressed with better education.   A more careful examination of methodologies is much more valuable than blaming the use of $p$~values (\cite{leekpengnature}; \cite{Benjamini}).
   
 {\it Using the $p$~Value as a Tool for Discovery.}  Particle physicists make extensive use of $p$~values in deciding whether to reject the null hypotheses and claim a discovery.  These $p$~values are denoted as $p_0$ (see Figure~\ref{fig:p0p1}), as they represent a probability under the curve for the probability density of the data statistic, as predicted by $H_0$.  To claim a specific discovery, it is also necessary to check that the data are consistent with the expectation from $H_1$. Often in searches for new phenomena, the data statistic used for calculating $p_0$ is the likelihood ratio for the two hypotheses, $H_0$ and $H_1$. This already takes note of the alternative hypothesis. The cut-off on  $p_0$ is conventionally taken as $2.87\times 10^{-7}$, corresponding to a $z$-score of 5.0. \footnote{The $z$-score is the number of standard deviations from the central value of a standard normal distribution for which $p$ is fractional area under the portion of the curve for values greater than $z$; it is given by $z=\sqrt{2}\,\,{\rm{erf}}^{-1}(1-2p) = -\Phi^{-1}(p)$. This does not imply that our data statistic is normally distributed. It is simply used in the transformation of $p$ to $z$.}
 
 Some statisticians have criticized this criterion, pointing out that probability distributions are not so well known in their extreme tails, especially when their shapes are determined by systematic effects. The reasons in favor of requiring $z\ge 5.0$ are:
 \begin{itemize}
 \item{ Claiming a fundamentally new effect has widespread repercussions, and can have very high publicity. Withdrawing a claim of discovery can be embarrassing for a collaboration, and, more specifically, for the discovery proponents.  Reputations and future credibility can be tarnished.  The large author lists on particle physics experiments may serve as one reason for the extreme conservativeness in EPP, at least in recent decades.  Many collaborators who worked hard on their experiment but not on a particular analysis will be interested that their work does not contribute to claims that are later shown to be false. } 
\item{Past experience shows that effects with $z$-scores of 3 and 4 have often not been replicated when more data are collected (\cite{franklin2013shifting}).}
\item{A stringent standard on the $z$-score reduces the number of false claims made when systematic uncertainties are underestimated. This argument is most definitely not foolproof, as a systematic error that is not covered by an appropriate uncertainty can result in a false discovery of an effect with an arbitrarily high significance.  The five-sigma criterion effectively removes statistical fluctuations from the list of plausible explanations for a false discovery, focusing the discussion on systematic effects.}
\item{A five-standard-deviation effect is usually large enough that further studies can be done to investigate its properties.  Data can be divided into subsamples to see if the effect is present in all of the subsamples where it is expected, for example.  {\it In situ} studies of systematic effects have much clearer outcomes when the data sample is large enough.  It is possible to discover a particle or process with just one example interaction, however, if the predicted background is extremely low (\cite{Abbott:2016nmj}), hence the use of the word ``usually'' above.}
\item{It is important to have a pre-registered threshold for discovery that is shared by competing collaborations. Otherwise, credit for discoveries will go to those with the lowest standards.}
\item{The Look Elsewhere Effect can effectively increase a local $p$~value to a more relevant global $p$~value (see Section~\ref{sec:lee}).}
\item{An old but still relevant maxim is that ``Extraordinary claims require extraordinary evidence.'' Thus if we were looking for evidence of energy-nonconservation  in events at the LHC, we should require a $z$-score of much more than 5.0 before rushing into print. From a Bayesian viewpoint, this corresponds to assigning a much smaller prior probability to a hypothesis involving a radically new idea, as compared with traditional well-established physics.}
 \end{itemize}
 
The choice of $z\ge 5.0$ as the requirement for claiming a discovery is arbitrary, and it was made in order to realize the benefits listed above without suppressing an inordinate number of actual discoveries, at least given the characteristics of particle physics experiments.  It also encourages physicists to design experiments that are sensitive enough to unambiguously answer the questions they set out to address.  
 
 The very high standard for the use of the word ``discovery'' in EPP does not preclude the publication of results that provide indications that the null hypothesis may be false, but do not meet the five-sigma criterion.  Current convention stipulates that a result with a $z$~score of 3.0 or greater constitutes ``evidence'' in favor of $H_1$ over $H_0$.  Publications claiming evidence frequently contain full details of the experiment and the analysis,  ensuring that they are cited even after the discovery has been made. Often many improvements to an analysis are made after evidence has been achieved in order to raise the expected significance to that needed for discovery, hence papers announcing discoveries will highlight the changes to the analysis and the new data that have been added.
 
 While $p_0$ may be computed with Monte Carlo simulations of possible experimental outcomes, these calculations become very expensive with a threshold of $2.87\times 10^{-7}$.  Asymptotic formulas provide for more rapid calculation (\cite{Cowan:2010js}).
 Experiments often collect much more data than are needed to establish an effect with a $z$-score of 5.0.  Once the null hypothesis has been ruled out at high significance, the new effect becomes part of the null hypothesis for future tests of other alternative hypotheses, and parameter estimation becomes the most important result.  Particle physicists often say, ``Yesterday's sensation is today's calibration and tomorrow's background.''
 
 {\it Using $p$~Values to Reject Alternative Models.}  If we merely wish to exclude the alternative hypothesis, the convention is to use a $p$~value for $H_1$, denoted $p_1$ (see Figure~\ref{fig:p0p1}), of 0.05 or 0.10. This weaker requirement than that for rejecting $H_0$ is because the embarrassment of making a false exclusion is by no means as serious as that of an incorrect claim of some novel discovery.
     
An exclusion of models $H_1$ using the criterion $p_1< 0.05$ will falsely exclude $H_1$ at most 5\% of the time if $H_1$ is true.  However, most hypotheses of new physics are not true, and the risk of falsely excluding a true model is therefore rather low. 


Exclusions are usually expressed in terms of upper limits on signal strengths.  In 5\% of independent tests, assuming $H_0$ is true, all values of the signal strength including zero are excluded using this technique.  A plot of an upper limit on the signal strength as a function of a model parameter, such as the mass of a hypothetical particle, will then exclude 5\% of possible masses for all values of the signal strength, typically in disjoint subsets, assuming no new particle is truly present.  Physicists do not wish to exclude models that they did not test, even if their experiment's outcome is in what is called a ``recognizable subset''  (See \cite{mandelkern2002} and the comments by Gleser, Wasserman, van Dyk, Woodroofe and Zhang published alongside it for a lively discussion).  Furthermore, a plot showing exclusions all the way down to zero signal strength in 5\% of its tested parameters is not expected to be replicable -- the repeated experiment would have to get lucky or unlucky in the same way.
          
\begin{figure}[ht!]
\begin{center}
\includegraphics[width=0.7\textwidth]{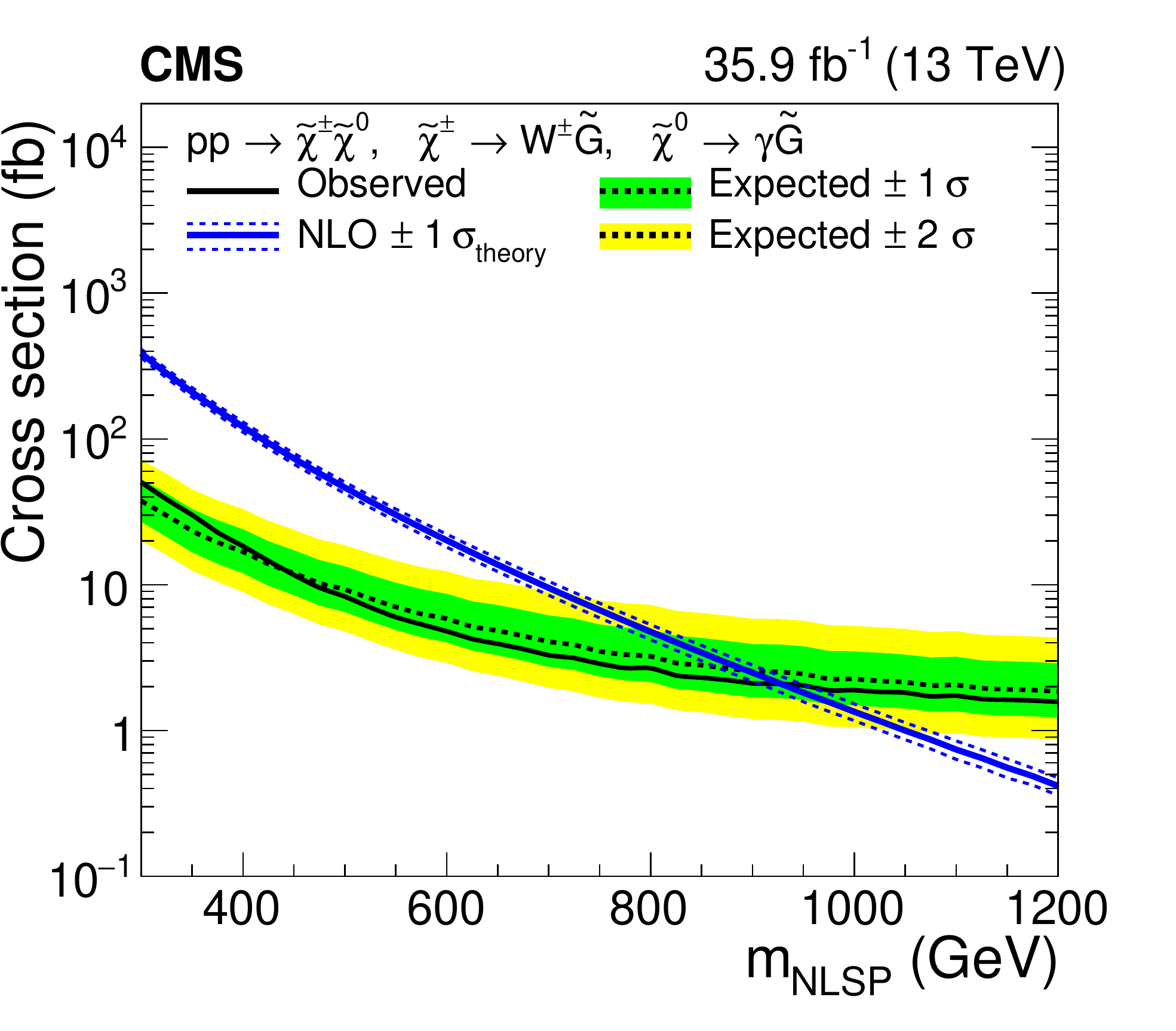}
\end{center}
\caption{Example of a ``limit plot'' from a search for supersymmetric particles from the CMS collaboration (\cite{Sirunyan:2018psa}).  The blue line shows the prediction of the production cross section in femtobarns (fb) from a specific supersymmetric model at next-to-leading-order (NLO) precision in the strong interaction, as a function of the mass of the next-to-lightest supersymmetric partner ($m_{\rm{NLSP}}$).  The black line shows the observed upper limit on the production cross section computed using the CL$_{\rm{s}}$ technique.  The dashed black line is the median upper bound calculated from a large sample of simulated repetitions of the experiment assuming the null hypothesis of no BSM particle production.  The green and yellow bands show the 68\% and 95\% intervals in which the limit is expected to vary, also assuming the null hypothesis.   The total integrated luminosity of the data sample is 35.9~fb$^{-1}$. With the assumptions of the model, particles with $m_{\rm{NLSP}}$ below 900 GeV are excluded, since for this mass range the upper limit on the production rate is below the one predicted by the model.}
\label{fig:limitplot}
\end{figure}

To combat the production of upper bounds on possible signal strengths being reported too small in 5\% of cases, particle physicists do one of two things. One option is to use a modified $p$~value, $p_1/(1-p_0)$ which is a ratio of two probabilities\footnote{The probability of obtaining exactly the test statistic value observed in the data is included in both the numerator $p_1$ and the denominator $(1-p_0)$.}, and which has been given the confusing name CL$_{\rm{s}}$ (\cite{Junk:1999kv}; \cite{Read:2002hq}).  If CL$_{\rm{s}}<0.05$ then $H_1$ is ruled out.  It has the property that CL$_{\rm{s}} \ge p_1$, and so comparing it with 0.05 will exclude $H_1$ no more often than the strictly frequentist test on $p_1$.  It also has the beneficial property of approaching 1.0 as the signal strength approaches zero, preventing exclusion of signals with zero strength.   An example of a plot of the upper limit on a production cross section as a function of a model parameter is given in Figure~\ref{fig:limitplot}.  If CL$_{\rm{s}}$ were not used and $p_1$ were used instead, the yellow band would reach down to zero and a random subset of mass points would be excluded with zero production rate. The other common technique is to use a Bayesian calculation of the posterior probability density as a function of the signal strength and exclude those values such that the integral from the upper limit to infinity of the posterior density is 0.05.   It also has the desirable property of producing credible upper limits that never are zero.

\subsection{The Look Elsewhere Effect (LEE)}
\label{sec:lee}
      
 Very often our alternative hypothesis is composite. If we are looking for a signal that produces a peak above a smooth background, the location of the peak, and also possibly its width, may be arbitrary. When we are  assessing the chance of random fluctuations resulting in a peak as significant as the one we see in our actual data, the ``local'' $p$~value is this probability for the given location in our data. Often, the smallest local $p$~value is the most exciting.  But more realistic is a ``global'' $p$~value, for having a significant fluctuation anywhere in the spectrum.   
It is the probability under the null hypothesis of observing anywhere in the spectrum a 
local $p$~value as small as or smaller than the lowest observed.  
This is similar to the statistical issue of  multiple testing, except that that considers discrete tests, while the LEE in the particle physics context often involves a continuous variable (e.g. the location of the peak).  The $p$~values at neighboring locations are often correlated due to detector resolution, so there are a finite number of independent tests even when the variable is continuous.   Asymptotic formulas exist to convert the local $p_0$~values to global ones, by taking into account the LEE in cases of a continuous variable such as the invariant mass of a possible new particle (\cite{Gross:2010qma}).
      
There are components to the LEE in addition to those described above. For example, the fluctuation could be in the distribution of the same physics variable, but varying the event selection requirements allows for fluctuations to appear or disappear. 
An increasingly common scenario is multiple teams analyzing the same data with different machine-learning discriminant variables.  A collaboration could choose to publish the results of the analysis with the lowest $p$~value, incurring a difficult-to-estimate LEE bias. Other data samples can be added providing more independent chances for a fluctuation.
These additional effects can be avoided by using a blind, or a pre-registered analysis, which cannot be tuned to produce a desired result.  If the choice of analysis or data selection requirements must be done after the data have been looked at, a good choice is to select the analysis that maximizes the {\bf expected} significance, or that minimizes the expected size of the interval on the parameter(s) of interest, without reference to the data outcome.  It is still possible to game this technique by considering a large number of adjustments to an analysis, each of which improves the sensitivity, and only including for publication those that make the observed result more like the desired result.  Careful review, including a discussion of conceivable analysis improvements, helps guard against these subtle effects.
      
Another complication is that the definition of ``elsewhere'' depends on who you are.  A graduate student might worry about possible fluctuations anywhere in his or her analysis, 
but the convenor of a physics group devoted to looking for evidence of the production of supersymmetric particles might well be worried about a statistical fluctuation in any of the many analyses searching for these particles.  In view of these ambiguities, it is recommended that when global $p$~values are being quoted, it is made clear what definition of ``elsewhere'' is being used. The definition of ``elsewhere'' may even change over time for a single analysis as more data are collected and the range of unexcluded model locations becomes smaller. 
It is also worth remembering that 
replicating a result that claims specific values of location parameters incurs at most a small look-elsewhere effect.  
      
Benjamini (\cite{Benjamini}) comments that in some cases non-replicability can be caused by the original analysis ignoring the effects of multiple testing. In a similar vein, an unscrupulous member of the news media or other interested reader may dredge the preprint servers for the most significant result of the month and not report all of the others that were passed over in the search.  Ignoring the LEE and reporting the smallest local $p$~value from a collection of them is a form of ``$p$-hacking'' (\cite{ioannidis2005}).
      
There is no LEE to take into account when computing model exclusions.  Each model parameter point is tested and excluded independently of others.  If a researcher sifts through all of the model exclusions looking for the most firmly excluded one and holds that up as an example, then an LEE may be necessary, but generally this is not of interest;  the set of excluded models is the important result.  In cases where multiple collaborations test the same model spaces and arrive at excluded regions of these spaces, then points in those spaces may have multiple opportunities to be falsely excluded.  The warning here goes to presentations of results in which excluded regions are merely overlaid on one another and the union of all excluded regions is inferred to be excluded.  In fact, a rigorous combination of the results is needed in order to make a single exclusion plot with proper coverage.

\section{Reproduction of Experimental Particle Physics  Results}
\label{sec:reproduction}

The definition of the word ``reproduce'' as used in this paper is the extraction of consistent results using the same data, methods, software and model assumptions (\cite{NAP25303}).  The only variables in this case are the human researchers, if there are any, and the separate runs on possibly different computers.  A failure to reproduce results can arise from improper packaging of digital artifacts, a lack of documentation, knowledge or even patience on the part of either the original researchers or those attempting the reproduction, the use of random numbers in the computational step, non-repeatability of calculations on computers either due to thread scheduling, radiological or cosmogenic interference with the computational equipment, or differences in the architectures of the computational equipment.  Section~\ref{sec:internalreproducibility} describes those issues of reproducibility that are encountered while a collaboration is producing its results, and Section~\ref{sec:reanarecast} describes data and analysis preservation, as well as the reuse of analyses to test models that were not considered in the original analysis, a procedure known as ``recasting'' a result.

\subsection{Ensuring Reproducibility as the Results are Made}
\label{sec:internalreproducibility}

In past decades, the internal representations of floating-point numbers in computers varied from one hardware vendor to the next, and results generally were not exactly reproducible when software was ported.  As late as the 1990's, physicists used mixtures of DEC VAX, IBM mainframes, Cray supercomputers and various RISC architectures, such as HP PA-RISC, IBM POWER, Sun Microsystems SPARC, DEC Alpha, and MIPS to name a few.  Many of these architectures had idiosyncratic handling of floating-point arithmetic.  Software had to be specially designed so that data files created on one computer architecture but read on another produced results as similar as possible.  

Large computing grids currently used in EPP contain mixtures of hardware from different vendors, though these are almost entirely composed of x86-64 processors manufactured by Intel and AMD. The relatively uniform landscape today makes reproducing results much easier, but by no means can differences in computer architecture be ignored.

Compiling programs with different optimization options can result in different results when run on the same computer, due to intermediate floating-point registers carrying higher precision than representations in memory.  If a program has an {\tt IF}-statement in it that tests whether or not a floating-point number is greater than or less than some threshold, a very small difference in a computed result can be the root cause of a very large difference in the rest of the output of a program.  EPP has long been sensitive to these issues, due to the complexity of the software in use and the large size of its data sets.  While each triggered readout of a detector is processed independently of all others, the large number of triggered readouts to process virtually guarantees that rare cases of calculations that perform differently, or even fail, will occur during a large data processing campaign.

Particle physicists have long recognized the utility of compiling software using different compilers, with all of the warnings enabled, and even with warnings treated as errors so they cannot be ignored during development.  Different compilers produce warning and error messages for different classes of errors in the source code, such as the use of uninitialized variables, some instances of which may go undetected by some compilers but which may be flagged by others.   Fixing software mistakes identified in this way helps guard against undefined behavior in programs.

Running the programs on computers of different architectures and comparing the results has also long been a tradition in EPP.  More recently, the process has been automated.  On each release of the software, or, in some cases, as fine-grained as on each change in any software component committed to a central repository, continuous integration systems now compile and link the software stack and run basic tests, comparing the results against previous results. These systems automatically warn the authors and the maintainers of the software of variations in the program's outputs.  These safeguards are very important in large collaborations because not every person developing software is an expert in every part of a large software stack, and unintended consequences of changes can go unnoticed or their root causes can be misassigned unless they are identified quickly.  Collaboration members who develop software sometimes leave for other jobs, placing high importance on good documentation and less reliance on individual memories of what the software does and how it works.  These continuous integration systems require human attention, as some changes to the software produce intended improvements, and these must be separately identified from undesired outcomes.

Reproduction of entire analyses is common in EPP when analysis tools are handed from one  analyzer, or team, to the next.  For example, when a graduate student graduates, a new student is often given the task of extending the previous student's work.  The first exercise for the new student however is to reproduce the earlier result performed by the previous student using the same data and software.  It is usually referred to as ``re-doing'' an analysis or ``checking'' an analysis.   Any discrepancy in this step points to a relatively simple flaw that can be remedied before proceeding.

Reproducibility is not a sufficient condition for reliability.  Reproducibility only tests the integrity of the computational steps, not the correctness of the assumptions entering the analysis or the quality of the input data.  A flawed result, when reproduced, contains the same flaws.

\subsection{Data and Analysis Preservation, Reuse and Recasting}

\label{sec:reanarecast}
Given the effort and expense of acquiring particle physics data, it is clearly mandatory for experimental
groups and collaborations to store their data and analysis assets in a way which makes it accessible for decades, in order for reproduction tests, replication tests, combinations with other results, and also to enable partially or entirely new analyses.  Vischia (\cite{VISCHIA2020100046}) provides example desiderata for data to accompany published results. 

{\it Strategies for Publishing and Preserving Analyses.} There are several approaches to address the preservation and the release of the data and analysis artifacts, with different audiences in mind.
\begin{itemize}
\item{Publish differential distributions of data and predictions.  This approach is the traditional one, in which histograms such as the one in Figure~\ref{fig:atlashzz4l} summarize the data in a manner that illustrates the main inputs to the statistical interpretation. Predicted yields and bin-by-bin contents can often be used to construct an approximate likelihood function and the distributions of its value under the hypotheses under test, though the uncertainties are often insufficiently parameterized.  A common failure of published physics results is the lack of communication of correlated uncertainties for the predictions in each bin of each histogram, which precludes the exact reproduction of the interpretation, such as a maximum likelihood fit. }
\item{Publish the likelihood function as a function of the parameter(s) of interest and all nuisance parameters.  This was proposed at the 2000 Workshop on Confidence Limits (\cite{James:2000et}) but only recently have experimental collaborations started to follow through on this proposal, notably from the work of Cranmer and collaborators on ATLAS (\cite{ATLAS:2019oik}).   Published analyses with likelihood functions on HEPData are available (\cite{Aad:2019vvf}; \cite{Aad:2019vvi}; \cite{Aad:2019byo}; \cite{Aad:2019pfy}).  With this information in hand, statistical interpretations following the likelihood principle such as Bayesian analyses can be applied by an outside party to reproduce the results.  Using a published likelihood to reproduce a result only tests some of the computational steps between the data and the result.  The steps needed to create the likelihood function are not reproduced but rather are reused.  A published likelihood can also be used to interpret results in the context of new models, provided that the new models' predictions are also predicted by the model for the published likelihood for some mapping of the parameters.}
\item{Publish the data, the simulated data, and the analysis code.  This is the most complete option, and it is starting to gain support given modern technology.  Distributing software in containers allows for reproduction of the analysis environment without concern for operating system dependencies.  Systematic uncertainties that have been evaluated based on the judgments made by the collaboration need to be encoded in the preserved and published software and data.}
\end{itemize}

The end users of publicly available data sets and analysis software have a variety of goals.  Some end users may wish to check a result and thereby learn how it is done so that they can apply this knowledge in their own work, either in a replication of the published analysis or to extend the techniques to other domains.  Some end users may wish to validate and/or tune the parameters of Monte Carlo simulation programs, either for the physics processes under study (\cite{Skands:2014pea}; \cite{Khachatryan:2015pea}; \cite{Sirunyan:2020pqv}) or for detector simulations (\cite{Elvira:2019fmq}; \cite{Battistoni:2015epi}).  Other end users may search for flaws in a published analysis, especially if they are trying to reconcile it with conflicting results.  The resolution of the conflict may involve updating model assumptions -- theoretical predictions, or simply the values chosen for nuisance parameters.  The places in the analysis at which they enter may be far upstream of the final results, requiring detailed raw information to include properly.  Similarly, a measurement may be improved if its nuisance parameters become better constrained after publication.  Still other end users may be interested in testing for new particles and processes that were not considered in the original publication and which may only have been thought of subsequently.  

{\it Recasting Analyses.}  Using published results to test models that are similar to but not the same as the models studied in the original publication is called ``recasting'' the results.  Recasting a result using the same model as the original result is a reproduction of the original result and is an important test of the completeness and the faithfulness of the preserved artifacts.

Recasting published results superficially resembles Hypothesizing After the Results are Known, or HARKing (\cite{doi:10.1207/s15327957pspr0203}).  An important distinction however must be drawn.  In an inappropriate HARKing process, post hoc hypotheses are presented as if they were a priori hypotheses, and the real a priori hypotheses are committed to a file drawer.  When an analysis is recast, however, the initial publication remains valid, and the recast result must be evaluated on its own merits.  An empty-handed search for a new particle or process recast as an empty-handed search for a similar particle or process can add value and knowledge from the same data.  A discovery that is recast in an alternative model asks a particularly valuable question.  Since a major part of a discovery is the exclusion of the null hypothesis rather than a confirmation of a specific alternative hypothesis, it is important to enumerate possible explanations for the deviation from the null hypothesis.  One of the possible pitfalls of a recast analysis is the expansion of the domain of ``elsewhere'' in the Look-Elsewhere Effect.  

Recasting an analysis that already has a full chain of data selection, discriminant calculation, and standard model predictions with uncertainties, still requires the selection efficiency and kinematic distributions of the new process to be modeled.  This step ideally requires full detector simulation but it may be performed with efficiency maps.
Tools have been developed to automate the reanalysis and recasting of EPP results (\cite{Cranmer:2018cqv}, \cite{Cranmer:2017frf}).  Analyses have already been performed using these tools (\cite{ATLAS:2020viz}; \cite{ATLAS:2019ivx}).   

A useful tool for abstracting and applying experimental event selection and classification criteria is Rivet (\cite{Buckley:2010ar}; \cite{Bierlich:2019rhm}).  To use Rivet, experimentalists provide implementations of their analysis logic as C++ plugin libraries, which are stored in a public repository (\cite{RIVETURL}).  These libraries can be used to produce histograms and other results using arbitrary data and Monte Carlo samples.  Recent developments in Rivet include tools for handling systematic uncertainties and detector emulation.  The LHC BSM Reinterpretation Forum provides a summary of available tools in use by theorists and experimentalists, and a set of recommendations for improving the presentation of results for future reinterpretation (\cite{10.21468/SciPostPhys.9.2.022}).

{\it Open Data.}  CERN has initiated both the Open Data and the CERN Analysis Preservation (CAP) projects (\cite{Chen2019}; \cite{DataPreservationCERN})
for storing the data and also all the relevant information, software and tools needed to preserve an analysis at the large experiments at the LHC. 
The preserved analysis assets include any useful metadata to allow understanding of the analysis workflow, related code, systematic uncertainties,  statistics procedures, meaningful keywords to assure the analysis is easily found, etc., as well as links to publications and to back-up material. 
This is a very involved procedure, and is still in its testing stage,  but it will clearly be very helpful for any subsequent reproducibility or replicability studies of the results of analyses using LHC data. 
It will, however, require a cultural change, with attention and effort on the part of physicists
performing analyses.

Initially, access to the stored information would be restricted to members of the collaboration who produced the data, but eventually it could be used by other EPP physicists and the wider range of scientists and the general public, on a time scale decided by the collaboration. 

Although developed at CERN for the EPP community, the CAP framework and its concepts may well be of interest to a wider range of scientists.  

{\it Barriers to External Reproduction of Analyses.} While the availability of data, models and analysis code has improved markedly over the past decade, the obstacles that have prevented it from happening earlier still exist and some must be overcome for each new analysis and data sample.  These include:
\begin{itemize}
    \item Analysis complexity.  Particle physics is an example of a field in which there are many steps that must be applied to data in order to extract a result:  reconstruction, calibration, event selection, event classification and statistical interpretation.  It is ambiguous how many of these steps must be redone in order to claim a successful reproduction of a result.  Usually the calibration and reconstruction steps are so cumbersome that they are done only once and the results are preserved and shared, even within a collaboration.  A collaboration may release data that pass certain selection requirements, but the decision to include an event in an analysis or exclude it may need to be reproduced, requiring all data, or at least data with loosened selection requirements, to be provided.
    \item Detector complexity.  An EPP collider detector contains many different components, not all of which may be functioning perfectly all the time, adding to the complexity of the analysis. Particles travel from one component to the next and the signals they produce must be associated with each other, either during reconstruction or event classification. Data must therefore be provided from all detector components for each event and not in separately-analyzable subsets.  Even if this association is already done in the prepared data, features of the components, such as angle and momentum coverage, affect the distributions the end user sees.
    \item Effort required from experts.  Data, analysis artifacts and documentation must be prepared so that non-collaborators can use them.  External users often require additional support in order to learn how to access the data and navigate unusual features.  Small collaborations may lack the personnel to support this, and even in large collaborations, support from funding agencies for data and analysis preservation is needed in order for the effort to be effective.
    \item Need for collaboration approval.  Reinterpretation of experimental data can easily produce incorrect results.  Many of the procedures described in Section~\ref{sec:reliability} depend on the review of the analyses and the associated uncertainties by the experimental collaboration from which the data come.  Simple mistakes, such as not applying a calibration factor, or more complex ones, such as adjusting data selection requirements after looking at the data, are often identified in internal review, which may be absent when data are analyzed by outside parties.
\end{itemize}
Given the increasing interest in the community for open data and analysis availability, and the availability of experts and resources for making it possible, these barriers are being overcome.  Modern tools allow better sharing of digital artifacts and detailed documentation helps address the support and quality-control concerns, but constant effort must be applied to help external parties use the preserved data and analysis artifacts of past analyses, and to preserve new ones.

\section{Replication of Experimental Particle Physics Results}
\label{sec:replication}

In order to gain confidence in the correctness of a result, scientists will perform a similar analysis, with the data and/or the analysis technique being different from the original ones.  The word ``replicate'' is defined to mean these kinds of independent tests (\cite{NAP25303}).  In EPP, however, the word ``replicate'' is rarely used in this sense due to the use of the word ``replica'' to mean an identical copy, as in data sets distributed to distant computer centers or in geometry descriptions of repeated, identical detector components.  Instead, ``independent confirmation'' is a more conventional phrase used in the case of successful replication, and ``ruling out,'' ``exclusion,'' and ``refutation'' are words that are used when replication attempts fail to confirm the earlier result.  If the data sets used in a replicated analysis overlap with those of the original analysis, the word ``independent'' is not used.  Replicated analyses often share sources of systematic error and thus also may fail to be independent even when the data sets, the experimental apparatus, and the collaborations are independent.  A common example of a replication of a result in EPP is to run the experiment longer, collect more data, and perform the analysis on the new data.  The statistical variations are expected to be different in the replica and the original, but the sources of systematic error are expected to be unchanged.

Both the measured values and the associated sensitivities are important to compare between an experiment and an attempt at replication.   A faithful replica of an experiment is expected to have a very similar sensitivity, but the actual data values may differ due to statistical fluctuations.  Using the same method to derive intervals when replicating an experiment and using compatible estimates of systematic uncertainties make the comparison simpler and less dependent on simulation.

In order to tell if a second experimental result successfully replicates the first, shared and independent sources of error must be carefully taken into account.  These are usually obtained from the quoted uncertainties on the measured values, but in the case of overlapping data sets, a component of the statistical uncertainty is also correlated.  The end result of a comparison of a replicated measurement is often a $p$~value expressing the probability that the two results would differ by as much as they were observed to or more, maximized over model parameters.  The sample space in which the $p$~value is computed consists of imaginary repetitions of the two experiments, assuming their outcomes are predicted by the same model.  A result with a very large systematic uncertainty is consistent with more possible true values of the parameter(s) of interest and thus passes replication tests more easily than one with a smaller systematic uncertainty.  Such a result is also less interesting because of its lack of constraint on the parameters of interest.

If a replication attempt fails, physicists are confronted with several questions: is one of the results incorrect?  Are multiple results flawed?  Or is there some explanation that can accommodate all of the results?  The last question, if answered in the affirmative, can prompt a great stride in the understanding of physics, or at least the measurements.  A replication attempt that is not an identical repetition the original experiment, but which rather changes some aspect of it, can be more informative than an exact repetition, as there may be questions that can be answered by examining the differences between the experiments.

\section{Strategies that Enhance the Reliability of Particle Physics Results}
\label{sec:reliability}

\subsection{Monitoring and Recording Experimental Conditions}
\label{sec:dataquality}

The conditions under which particle detectors are operated are known to affect their performance and thus may bias the results obtained from them.  Particle physics experiments run for years at a time, and operating conditions are variable, making the data sets heterogeneous.  Variations in accelerator parameters, such as the beam energy, the energy spread, the intensity and stray particles accompanying the beam (``halo'') are constantly monitored and automatically recorded in databases for future retrieval during data analysis.  Environmental variables such as ambient temperature, pressure and humidity are also included in these records.  The concentrations of electronegative impurities in drift media are constantly measured and recorded.  The status of high-voltage settings, electronics noise, and which detector components are functioning or broken are also recorded.  Non-functioning detector components are often repaired during scheduled accelerator downtime, and some detector components may recover functionality when computer processes are restarted.  If a particular physics analysis requires that specific parts of the detector are fully functional, then only data that were taken while the detector satisfies the relevant requirements can be included in that analysis.

While an experiment is collecting data, physicists take shifts operating the detector and monitoring the data that come out of it.  While most detector parameters that affect analyses can be identified in advance and monitored automatically, some surprises can and do occur. It is up to the shift crew to identify those conditions, notify experts who may be able to repair the errant condition, and mark the data appropriately so they do not bias physics results.  The shift crew is aided by automated processes that analyze basic quantities of the data in near real time, providing input to their decisions.

\subsection{Software Version Control}
\label{sec:softwareversioncontrol}

Because the data processing and analyses in EPP require the use of a large amount of software that is under constant development by a large number of people, a well thought-out version control system is required.  Not only must the source code be under strict version control, but so too must the installed environments, which include auxiliary files and databases.  Naive systems in which collaborators share computers on which the software is constantly updated to the latest version will find their analysis work difficult in a way that scales with the size and activity of the software development effort.  Results obtained by a physicist running the same programs on the same data may differ from day to day, or programs that ran previously may fail to run at all.  A new release of a software component may be objectively better than the older ones -- bugs may have been fixed or the performance of the algorithms may have been improved.  ``Performance'' here refers not to the speed with which the program runs on a computer, but rather to its ability to do its intended job.  An improved track-reconstruction algorithm may correctly reconstruct more tracks or find fewer false tracks than an older version, for example.  But if an analyzer has measured the performance of the algorithms using experimental data if possible and Monte Carlo simulations otherwise, then those algorithms must be held constant or the calibration constants become invalid.  One may ``freeze'' the software releases, but then some collaborators will require newer releases than other collaborators.

The solution chosen in EPP is to freeze and distribute pre-compiled binaries and associated data and configuration files for each release used by any collaborator.  No software version is set up by default when a user logs in -- a specific version must be specified.  The version for a top-level software component determines the versions needed for all dependent components, which are automatically selected.  Inconsistent version requests are treated as errors.  New users are surprised at the need for this complexity, but they appreciate it later when they are finishing up their analysis work and are trying to keep every piece of their workflows stable.

\subsection{Reporting the Expected Sensitivity}
\label{sec:sensitivity}

A common and necessary practice in EPP is to compute the expected sensitivity of an analysis before the data are collected, or at least before they are analyzed.  Since particle physics experiments are so expensive and take so long to design, construct, operate, and perform data analysis, funding agencies require that a collaboration must demonstrate that their proposed experiment is capable of testing the desired hypotheses before approving the funding.  The expected sensitivity usually takes the form of the expected length of the confidence interval on one or more measured parameters, the median expected upper limit on the rate of a process assuming it truly is not present in Nature, or the median $p_0$~value, assuming the new process truly is present.  Distributions of possible outcomes of the experiment are pre-computed when an experiment is proposed.  After the experiment has run for a while, updated knowledge about nuisance parameters can be obtained by analyzing subsets of the data, and the sensitivities and the distributions of possible outcomes are recomputed before the observed result is obtained.   The observed result can then be compared with the expected results.  Sometimes a spurious outcome is easily identifiable as being consistent with none of the considered hypotheses.  In analyses that seek new particles or interactions, the separate calculation of sensitivity and significance helps combat the ``file-drawer'' effect.  Even if a result fails to be significant, if the sensitivity of the test is high, then the result is worth publishing. 
      
\subsection{Standard Candles and Control Samples}
\label{sec:controlsamples}
    
One way in which physics analyses can benefit from replication without waiting for another group on the same or a different collaboration to work on a similar analysis is to use a calibration source, or a ``standard candle.''  Signals that have been long established ought to be visible in analyses that seek similar but not-yet-established signals.  The analysis therefore replicates part of the earlier work, and in so doing, not only validates the earlier work, but increases the confidence in the present work.  This step is particularly important in analyses that do not observe a new signal.  One might think that the detector or the analysis method is simply not sensitive to the new signal and may have missed it.  To show that a known signal is found in the same analysis with the expected strength and properties gives confidence that the whole chain is working as desired.
    
A related technique is the use of ``control samples,'' in which the desired signal is known not to exist, or if it does, contributes a much smaller fraction of events than in the selected signal sample.  Subsets of the data that do not pass selection requirements, or that were collected in different accelerator conditions (off-resonance running is an example of a way to collect background-only data) can be used to estimate the rates and properties of processes that are not the intended signal but which can be confounded with it if not carefully controlled.  Often multiple control samples are used, each one targeting a specific background process, or which may constrain the rates or properties of them.  Disagreements in the predictions of background rates and properties from different control samples are often contributions to the systematic uncertainty estimations used in the signal sample.

\subsection{Dedicated Calibration Working Groups}
\label{sec:calibration}
    
Large EPP collaborations have a difficult, complex task to perform to produce any individual result.  The detectors have millions of active elements, and the conditions are variable.  Often a physics analysis requires events to be selected with a specific particle content -- say a charged lepton, a number of jets, and missing energy.   Not every lepton is identified correctly, however, and not every jet's energy is measured well.  Instead of requiring every team that wants to analyze data to perform the work to calibrate all of the things that need calibrating, working groups are set up to perform these tasks.  A group may be devoted just to $b$-jet tagging while another will calibrate the electron identification and energy scale.  Other groups will form around each necessary task.  Their results, along with systematic uncertainty estimates, are reviewed in much the same way as physics results are reviewed before being approved for use by the collaboration.  In this way, consistent calibrations are available for all physics analyses performed by the collaboration, and mistakes are minimized.  One avenue that subconscious bias can affect an analysis is in the calibration stage.  The separation of the calibration efforts into dedicated groups reduces the possibility that collaborators wishing specific results for their analyses can do so via (subconsciously) manipulating calibrations, as each calibration group must provide results for everyone in the collaboration, not just one set of interested parties.  

\subsection{Blind Analyses}
\label{sec:blind_analyses}

Blind analysis methods are designed to prevent experimenters from consciously or unconsciously manipulating their data analyses in order to obtain desired results.  An example of analysis manipulation is the adjustment of data selection requirements or the event classification procedure in order to maximize the size of the observed signal in the data.  While such adjustments may have the intended effect of improving the signal-to-background performance of the analysis, they also can emphasize statistical fluctuations by including only those regions in observed variable space that have excess events in them.  On average, analyses optimized in this way produce $p$~values that are too small and signal strength measurements that are too large.  If it is discovered during review that an analysis did not begin with well-defined event selection and classification procedures but rather it was optimized to maximize the result, the data used in this optimization are discarded for that analysis and additional data must be used.  To avoid situations like these and use all the data, analyses are blinded to the data from the beginning.

Several methods for blinding analyses have been and continue to be used in EPP (\cite{Klein:2005di}).  One simple method is to optimize the analysis using simulated data and reserve the input of data from the experimental apparatus until the procedures have been decided upon, including the data selection, classification, and statistical procedures.  Analyses are constructed and optimized based on predicted outcomes of expected signal and background contributions to the event yields, and so it is usually possible to perform the necessary steps without access to the data from the experiment.  The collaboration must then agree to accept the result of the analysis after the data are input to the analysis (``unblinding''), without change to any step of the analysis, or the procedure is not fully blind.

A drawback of the simple blinding procedure described above is that it precludes the use of data from the experimental apparatus as a calibration source to help constrain the values of  nuisance parameters and to help guide the data selection.  This shortcoming is addressed by partial blinding.  Data in control samples -- events that fail one or more signal selection requirements, for example, are allowed to be input to the analysis procedure before the selected ``signal'' sample is made available for analysis.  Sometimes surprises are found in the control samples -- previously unappreciated sources of background events or miscalibrations can show up at this stage.  The process of eventually revealing data passing signal selection requirements is often called ``opening the box.''  The process relies on the good faith of the collaboration members not to look at data that have been blinded.  In a large collaboration with many different analyses being developed in parallel, sometimes 
one analysis group's control sample is another group's selected signal sample.  However, now that sophisticated ML procedures are commonplace, one group's histogram of a highly-specific ML discriminant variable is unlikely to be meaningful to another group that may be seeking a different sort of signal entirely.
      
A similar strategy for partial blinding which helps prevent big surprises from showing up when the signal box is opened is to intentionally look at a small subset of the data
and to use this to tune the way the analysis will be performed. This is then frozen for use on the remainder of the still-blinded data.   The early data than was used to define the analysis must then be omitted from the final result if a fully blind analysis is to be claimed.  This procedure is the similar to the method described above to turn a manipulated analysis into a blind analysis, but planning in advance how much early data are to be used in a non-blind fashion allows the sensitivity of the analysis to be optimized when the total data set is limited in size.

Yet another blinding procedure which applies to precision measurement of physical quantities is to encode an arbitrary, fixed offset in the final step of parameter inference in software, and to hide the value of this offset from researchers performing the analysis work.  This offset is then removed before publication, in a step called ``unblinding the value.''  This method is not used for significance tests.

\subsection{Redundant Experiments}
\label{sec:redundantexperiments}
    
Circular colliders typically have multiple detectors located at discrete interaction regions that produce identical physics processes because the beams are the same.  The PEP ring at SLAC had five detectors: HRS, TPC-2$\gamma$, Mark-II, MAC and DELCO.  The LEP collider had the ALEPH, DELPHI, L3 and OPAL detectors.  The detectors at the LHC include ATLAS, CMS, ALICE and LHCb.  These detectors are only partially redundant; they are not exact copies of one another.  Part of the purpose is to provide for replication of results, but it is also important to diversify the technology used in the experimental apparatus.  While detector research and development is also a mature field and technologies are deployed in large experiments only if they have been shown to work in prototypes, risks still exist.  A particular technology may be more ideal for a specific physics analysis than another, but it may be weaker in a different analysis.  Given the high costs of these detectors, the additional value accrued by exposing different technologies to the same physics is seen to be a better investment than exact duplication.  Competition between collaborations also encourages scientists to optimize their analyses for sensitivity (and not significance), and to produce results quickly so as not to lose the race with competitors.  High-profile discoveries, such as those of the top quark (\cite{Abe:1995hr}; \cite{D0:1995jca}) and the Higgs boson  (\cite{Aad:2012tfa}; \cite{Chatrchyan:2012ufa}), are often simultaneously announced by rival collaborations using similarly-sized data sets.
    
\subsection{Testing the Consequences of a Result}
\label{sec:testingconsequences}

Results are frequently interpreted in the context of other results obtained in similar but not identical processes, but for which the model explanation for one experiment's result must have consequences for another experiment's result.  An example of this is the search for a fourth light neutrino with exotic properties.  The LSND and MiniBooNE collaborations observed  excesses of $\nu_e$ events in beams dominantly composed of $\nu_\mu$, when compared to what was expected given what is known from three-flavor neutrino oscillation rates (\cite{Aguilar:2001ty}; \cite{Aguilar-Arevalo:2013pmq}; \cite{Aguilar-Arevalo:2018gpe}). While $\nu_\mu$s are expected to oscillate into $\nu_e$s in the three-flavor model, the observed rate of appearance of $\nu_e$ is too large given the short distances the neutrinos traveled, the energies of the neutrinos, and the known oscillation parameters measured by other experiments.   A hypothesis to explain these data is that a fourth neutrino may exist which provides another oscillation path from $\nu_\mu$ to $\nu_e$, one that oscillates faster than the three-neutrino model prediction.  We know from the LEP experiments' $Z^0$ lineshape measurements (\cite{ALEPH:2005ab}) that there are only three light neutrino species that interact with the $Z^0$ boson.  A fourth light neutrino must therefore be ``sterile.''  Nonetheless, in order to explain the LSND excess in this way, some $\nu_\mu$s must disappear as they oscillate into sterile neutrinos, while a few of these sterile neutrinos may oscillate back into $\nu_e$.  One can test this hypothesis by looking for $\nu_\mu$ interactions in a $\nu_\mu$ beam (any $\nu_\mu$ beam, not necessarily LSND's or MiniBooNE's), and see if enough disappear as a function of the distance from the neutrino source divided by the neutrino energy.  Measurements can also be made of the fraction of $\nu_e$s that disappear in a sample of $\nu_e$s, or similarly with samples of antineutrinos. A recent combination of data from the MINOS, MINOS+, Daya Bay and Bugey experiments which have measured these disappearance rates (\cite{Adamson:2020jvo}) exclude parameter values consistent with the LSND result.  While the newer experiments did not attempt to directly replicate the LSND or the MiniBooNE experiments, they do constrain its interpretation.  It remains to be seen whether the tensions in this field come from inadequate understanding of experimental effects, or from a more fundamental physical process, even if it may not be a light, sterile neutrino.
    
\subsection{Review of Results}
\label{sec:review}
    
Large collaborations benefit from the availability of scientists with diverse experiences and points of view.  All collaborators on the author list are given the opportunity to review each result that is published.  While the number of papers published by each of the LHC collaborations is such that not every collaborator reads every paper, each collaborating institution is required to meet a quota of papers that are read and commented on by its members.  Below, we describe the review procedures used by CDF, a typical hadron collider collaboration.
    
{\it Preliminary Results.}  Results in preparation must pass through a lengthy, formal approval process before they can be presented outside of the collaboration.   Working groups led by experienced physicists review each analysis by the members of the group and frequently point out flaws in logic, data handling, analysis, documentation and presentation.  Before approval, a result must be fully presented to a working group, and a public note must be drafted.  A public note is a document that fully explains the analysis to an intended audience of physicists outside of the collaboration.  At this stage, members of the working group ask the proponents questions about their analysis and the documentation. Some of these questions may require significant additional study.  At a later date, the analysis must be presented again, and all questions and requests must be answered to the satisfaction of the group members.  Only at this phase can a result be approved, though all figures and numbers must be labeled ``Preliminary.''  The public note is made available on the collaboration's physics results web page, and the result may be presented at conferences and seminars.
    
Preliminary results sometimes do not have the final estimates of systematic uncertainties associated with them.   Even the list of all potential sources of systematic error may not be fully understood at the time a preliminary result is reviewed.  An incomplete accounting of systematic uncertainties is likely to result in an underestimate of the total uncertainty.  If a result needs to be produced on a short timescale under these circumstances, systematic uncertainties are estimated conservatively -- overestimates are preferred to underestimates. The intention is that further work will reduce their magnitudes (\cite{Barlow:2002yb}).
    
There are negative consequences to overestimating systematic uncertainties.  A set of measurements of the same physical quantity by several collaborations, some or all of which overestimate their uncertainties, will have a $\chi^2$ that is smaller than expected, even if the sources of systematic error are different.   More worrisome, however, is the possibility that a combination of results may assume a measurement is more sensitive to a particular nuisance parameter than it really is, due to an inflated or misclassified source of uncertainty.  The measurement thus serves to constrain the nuisance parameter too strongly in the joint result, sometimes producing a final combined result with an underestimated uncertainty.  
    
{\it Publication.}  After  a preliminary result is released, the physicists who performed the analysis prepare a manuscript for publication.  At this stage, and sometimes even during the preliminary result preparation stage, a committee of collaborators who have worked on similar topics but who are not directly involved in the particular result is set up to review the paper draft.  Often the committee is involved at an early stage of writing the draft and they meet regularly with the authors to improve the analysis and the presentation. All changes to the analysis must be approved by the  working group specializing in the topic.  By the time a manuscript is released to the collaboration for review, the result has already been reviewed multiple times.  The additional scrutiny from the large collaboration may uncover additional flaws, and the manuscript is re-released for a second collaboration review after addressing the concerns raised in the first review.  The process is iterated until consensus is reached that the paper can be submitted for publication.  This process often uncovers even tiny flaws and it can take several months or even years to complete. High-profile results can be pushed through on accelerated timescales without compromising the integrity of the review, provided that the necessary effort can be directed towards them.   Only in very rare instances is consensus not reached.  In these cases, dissenting collaborators can request their names to be removed from the author list of that paper.  A significant fraction of the collaboration refusing to sign a paper provides a strong signal to the analysis proponents and the readers of the article about the perceived validity of the results.  
    
After a manuscript has been agreed upon and submitted to a journal, the editors use traditional blind peer review before publication.  Referees who are known to be experts in the field, frequently who are also members of rival collaborations, weigh in on the publication.  Referees are not superhuman, though they do sometimes find issues with papers that thousands of authors may have missed.  Sometimes a collaboration may fall into the trap of ``group-think,'' having repeated the same arguments to itself over and over again, so an independent check has as much value in EPP as in other fields.  An independent review also can help improve the presentation of work that may not be clear to an outsider.  Sometimes the root cause of the non-replicability of a result is merely inadequate or unclear documentation.
     
{\it Statistics Committees.}  Large collaborations sometimes have a ``statistics committee,'' which is made up of collaborators who are experts on data analysis, inference, and presentation of results.  A good-sized committee has at least six members, and more are desirable.  Statistical issues in analyses can be intricate and they take some time to understand. Members of the committee sometimes disagree about issues with specific analyses.  It is important for physicists who are embarking on a new analysis to consult with the collaboration's statistics committee, so that work is not steered in a direction that is only later found to be flawed under collaboration review.  Frequently, the most challenging issues with an analysis relate to the treatment of systematic uncertainties.  The enumeration of the sources of uncertainty, their prior constraints, how to constrain them {\it in situ} with the data, and how to include their effects in the final results, are common subjects that the statistics committee must address.  Experimental collaborations typically must each have their own statistics committee, as results in preparation are usually confidential until a preliminary result is released, and review by members of other collaborations would spoil this confidentiality.  Peer review generally does not spoil confidentiality because all or nearly all results are submitted as preliminary results first, and preprints are available for submitted manuscripts, thus establishing priority.  Members of a collaboration may be wary of advice from members from competing collaborations, even if it is general statistical methods advice.  Members of one collaboration may prefer that their competitors treat their uncertainties more conservatively and thus appear to have a less reliable result.  Even the fear of such bias in advice is enough to prevent the formation of joint statistics committees across collaborations.
    
Munaf\`o and collaborators (\cite{munafo2017}) point out that  independent methodological support committees have been very useful in clinical trials.  Particle physicists routinely reach out to statisticians, holding workshops titled PhyStat every couple of years (\cite{phystat}).
    
{\it Combining Data from Multiple Collaborations.}  Often it is useful to produce more sensitive results by combining the data\footnote{Combining the data is in general better (but more complicated) than simply combining the results.} from competing collaborations.  These combined results are sometimes presented at the same time as the separate results.  In this case, the collaborations must agree on the exchange of data and appropriate methods of inferring results.  The methods used typically are extensions of what the collaborations use to prepare their own results, and usually in a combination effort, members of each experiment perform the combination with their own methods and the results are compared for consistency.  Much of the discussion between collaborations working to produce a combined result centers on the identification, the magnitudes, and the procedures for handling systematic uncertainties.  During combination, mistakes in the creation or the exchange of digital artifacts may be exposed and they must be addressed before the final results can be approved by all collaborations contributing to the combined results. 
     

\section{ Special Features of Particle Physics Results}
\label{sec:specialfeatures}

Nearly all results in EPP are derived from counts of interactions in particle detectors.  Each interaction has measurable properties and may differ in many ways from other interactions.  These counts are typically binned in histograms, where each bin collects events with similar values of some observable quantity, like a reconstructed invariant mass.  An example of such a histogram, showing the distribution of the reconstructed mass $m_{4\ell}$ in $H\rightarrow ZZ\rightarrow 4\ell$ decays, selected by the ATLAS collaboration (\cite{ATLAS:2020wny}), is given in Figure~\ref{fig:atlashzz4l}.
Event counts often are simply reported by themselves.  Under imagined identical repetitions of the experiment, the event counts in each bin of each histogram are expected to be Poisson distributed, although the means are usually unknown or not perfectly known.  The data provide an estimate of the Poisson mean, which is often directly related to a parameter of interest, such as an interaction probability.    Data in EPP have been referred to as ``marked Poisson'' data, where the marks are the quantities measured by a particle detector, such as the energies, momenta, positions and angles of particles produced in collisions.  The fact that all practitioners use the same underlying Poisson model for the data helps reproducibility and replication.

An advantage of particle physics analyses as compared with those in, for example, sociology is that the behavior of elementary particles that are the subjects of the study are affected by few to no confounding variables and thus can be relied upon to be repeatable.  There is no need to worry whether a sample of muons is representative of muons in general with respect to age, salary and socio-economic class, nor whether they are informed about their treatment during the experiment.  
``Double-blind'' analyses do not exist in particle physics.  The blind analysis procedures described in Section~\ref{sec:blind_analyses} refer to blinding the researchers to the results in preparation. 
Aside from the randomness arising from quantum mechanics, particle interactions can be controlled by preparing their initial states in particle accelerators.  The probability distributions arising from quantum-mechanical randomness are well-defined and repeatable, and they are often what is being measured.

Another difference between particle physics and other areas of research is the way models are regarded. The basic model used for particle physics is the standard model, which contains the particles listed  in Table~\ref{Table:generations}, and the forces in Table~\ref{Table:Forces} (apart from gravitation, which for most purposes can be neglected on a particle scale). 
The standard model provides an excellent description of a large number of experimental distributions, but despite this it is not believed to be the ultimate description of Nature; for example, it does not explain dark matter or dark energy, and it has about 20 arbitrary parameters (such as particle masses) that have to be determined from experiment rather than being predicted by the theory. Thus we have a model that works very well, but which we are constantly hoping to disprove. Most of our ``search'' experiments are seeking not merely to produce another verification of the SM, but are hoping to discover evidence for physics beyond the SM.
A convincing rejection of the SM would be a major discovery.

The different goals of research in different fields affects the cost-benefit calculation of the desired level of significance required to make a discovery. 
Campbell and Gustafson (\cite{doi:10.1080/00031305.2018.1555101}) review proposals to lower $\alpha$ from its common value of 0.05 in non-EPP fields of science.  They investigate the impacts such proposals on the Type-I error rate, the publication rate, and the rate of suppression of actual discoveries.  Particle physics experiments have very different costs and benefits compared with drug trials, for example.  Truths about fundamental interactions between matter and energy will stay true even if a discovery is delayed while more data are collected, while in a drug trial, the lives and the health of people are affected by delays.  The use of the particles in the study is not subject to ethical concerns as are human subjects.  Most of the cost of an experiment is in the construction of the apparatus, though the cost of operations over many years may also be quite large.  For this reason, it usually makes sense to run particle physics experiments around the clock for as long as the expected significance of potential discoveries continues to rise quickly enough, and/or the expected size of confidence intervals on parameters of interest to fall quickly enough, to justify the running expenses.  When the main physics goals of an EPP experiment become limited by systematic uncertainty, or if a much better apparatus is built elsewhere, operations cease.  For these reasons, the $z\ge 5.0$ criterion for discovery makes more sense in EPP than it does in medical trials, for example.

Another peculiarity of EPP is the specialization of practitioners into theoretical and experimental categories.  There are two important benefits to this division.  Models to be tested generally are published in theory or phenomenology papers before the experimental studies are conducted.  This serves to fully define the models before they are tested, though many models have a finite number of adjustable parameters that experimentalists must consider.  The second benefit is that experimentalists almost never test theories that they themselves invented, helping to reduce possible effects of confirmation bias.

Furthermore, while there is only one true set (however incompletely known) of physical laws governing Nature, the set of speculative possibilities is limited only by physicists' imaginations.  Theory and phenomenology preprints and publications abound in great numbers.  Most theories are in fact not true, but generally in order to be published, they must be consistent with existing data.  Experimenters are well aware that most searches for new particles or interactions will come up empty-handed, even though the hope is that if one of them makes a discovery, then our understanding of fundamental physics will make a great stride forwards.  Null results can also contribute valuable scientific knowledge.

The LHC has been referred to as a ``theory assassin,'' owing to all of the null results excluding many speculative ideas.  This process largely mitigates the ``file-drawer'' effect, as a null result excluding a published theoretical model is likely to be publishable and not ignored.  The experimental tests must be accompanied with proof that they are sensitive to the predictions of the theories in question, however, before they are taken seriously for publication.  Null results are therefore typically presented as upper limits on signal strengths.  Theorists may counter a null result by predicting smaller signal strengths or effects that may have evaded the experimental tests in other ways.  Proposed new particles may be too massive to produce in collisions at the accelerator at which the limits were set, or they may decay in ways that are difficult to distinguish from background interactions, for example.   Frequently, an iterative process is undertaken that progressively tightens the constraints on a model of new physics as more data are collected and analysis techniques are improved.  Some models can never be fully ruled out because signal strengths could always be smaller.  If the unexcluded parameter space of a model is  interesting enough to constrain further or to motivate a possible discovery,  it is left as a challenge to the next generation of experiments.  New apparatus may be required, such as higher-energy accelerators or more sensitive detectors.  Novel approaches, such as using natural sources, e.g. cosmic rays, or asking questions that may be answered by astrophysics or cosmology, may be brought to bear.

\section{Examples of Non-Replicated Results in Particle Physics}
\label{sec:examples_nonrep}
       
       \subsection{Free quarks}
       \label{sec:Quarks}
The success of the quark model in explaining the spectroscopy of the many existing hadrons (and the absence of those that were forbidden by the model), as well as many features of the production processes for interactions, resulted in many searches for evidence of the existence of free quarks. Most of these used the fact that quarks have an electric charge of $\pm e/3$ or $\pm 2e/3$. Experiments looked for quarks in cosmic rays, reactions at accelerators, the Sun, moon dust, meteorites, ocean sludge, mountain lava, lobster shells, etc., but almost all experiments yielded null results. Theorists accepted this as being due to the concept of ``confinement'': quarks can exist only inside hadrons, but not as free particles. 

However, in 1981, an experiment at Stanford reported a positive result (\cite{Larue:1981jc}).
It involved levitating small spherical niobium spheres, and measuring their oscillations in an oscillating electric field; their amplitude is proportional to the charge on the ball. Of 39 measurements reported, 14 corresponded to the fractional charge of quarks. Although the word ``quark'' does not appear in the Stanford publication,  
this was possible evidence for their existence as free particles.

However, many analysis decisions had to be made in order to extract the charge on a ball from the raw measurements e.g. whether or not to accept an experimental run, which corrections to apply for experimental features, etc. These decisions were made while looking at the possible result of the charge measurement. Luis Alvarez suggested that a form of blind analysis should be used (see Section~\ref{sec:blind_analyses}). This involved the computer analyzing the data adding a random number onto the extracted charge visible to the physicists; this was to be subtracted from the result only after they had made all necessary decisions about run acceptance and corrections. The consequence was that this experiment published no further results. 
       

\subsection{Additional examples of non-replicated results}
\label{sec:additionalexamplesnonrep}
       
In his review, ``Pathological Science,'' Stone (\cite{Stone:2000an}) collects several stories of experimental results that have later been found to be wrong.  In several of his historical examples, flawed methodology was uncovered during visits by experts to the laboratories where the spurious discoveries had been made.  Franklin (\cite{10.1088/978-1-64327-162-0}) provides excellent commentary on some results that have been successfully replicated and some that have not.  Bailey (\cite{Bailey:2016lva}) has produced histograms of changes in measured values of particle properties expressed in terms of the reported uncertainties.  While many repeated measurements of the same quantities are consistent, there is a long tail of highly discrepant results.

\section{Examples of Over-Replicated Results in Particle Physics}
\label{sec:example_toomuchrep}

More concerning than false results that are not replicated are false results that are replicated.  Of course, these results can only be ascertained as false by further attempts at replication and/or the discovery of errors in the original results.


\subsection{Pentaquarks}
\label{sec:pentaquarks}

The search for a pentaquark is an interesting example of replication. Particles known as hadrons are divided into baryons and mesons. In the original quark model, baryons are composed of three quarks (and mesons of a quark and an antiquark)- see Section \ref{sec:WiPP}. There was, however, no obvious reason why baryons could not be made of four quarks and an anti-quark; these baryons would be pentaquark states. This would make available new types of baryons which could not be made of the simple and more restrictive three-quark structure, and which could be identified by their decays modes involving unconventional groupings of particles not accessible to three-quark baryons. Thus searches were made for these new possible particles. In 2003, four experiments provided evidence suggesting the existence of one of these possibilities, known as the $\Theta^+$, with a mass around 1.54~GeV.  The quoted significances were 4 to 5 $\sigma$. Indeed, national prizes were awarded to physicists involved in these experiments. In the following couple of years there were six more experiments quoting evidence in favor of its existence.

However, other studies, many with much higher event numbers than those with positive results, saw no evidence for the particle. Although most of these were not exact replications of the original positive ones, at least one was a continuation with much higher event numbers than the original study, and involved exactly the same reaction and the same beam energy; it failed to confirm its original result. 

The net conclusion was that the $\Theta^+$ does not exist. Possible reasons for the apparently spurious early results include poor estimates of background, non-optimal methods of assessing significance, the effect of using non-blind methods for selecting the event sample and for the mass location of the $\Theta^+$, and unlucky statistical fluctuations.   Hicks (\cite{Hicks:2012zz}) provides a detailed review of pentaquark search experiments and their methodologies.

This topic is probably the one in which there were the most positive replications of the discovery of a particle that does not exist. It demonstrates the care needed when taking a confirmatory replication as evidence that the analyses are correct, especially when the experiments involve smallish numbers of events.

The twist in the tale of this topic is that more recently, pentaquark states have been observed by the LHCb experiment (\cite{Aaij:2015tga}). They are, however,  much higher in mass than the $\Theta^+$, and have a different quark composition, so are certainly not the same particle. More details of the interesting history of the search for pentaquarks and their eventual discovery and measurement can be found in the review on the subject by M.~Karliner and T.~Skwarnicki in the 2020 Review of Particle Physics (\cite{Zyla:2020zbs}). 

\subsection{Parameter Determination}
\label{sec:parameter_determination}

It is not only searches for new particles that can suffer from spurious replication, but measured values of well-established particles and processes can also be affected.  The Particle Data Group collects measurements of particle properties, averages them in cases of multiple measurements of the same quantity, and publishes these every two years (\cite{Zyla:2020zbs}). One can see in the historical evolution of the averages that the error bars generally decrease over time and the differences between the measured values also decrease over time.  There is considerable correlation from one average to the next, which is largely due to the same measurements contributing to multiple years' averages.  In order to see if there is an effect in which experimenters seek out, consciously or not, to replicate earlier numbers without contradicting them, a meta-analysis was performed (\cite{Klein:2005di}) in which individual measurements of selected quantities were plotted as functions of time.  Correlations are indeed visible even in these historical plots.  Not all of the effects may be due to over-eagerness to replicate earlier work, because often shared sources of systematic error afflict multiple measurements.

\section{Conclusions}
\label{sec:conclusion}


Experimental particle physicists publish reliable results nearly all of the time.   Many factors contribute to this.  The models being tested are almost always well defined in advance of the experimental test. High standards are applied to the results.  Data processing, calibration, analysis and review tasks are divided among many collaborators.  Internal cross-checks of the results are possible in many analyses and are used when available. Many analyses are blind. Stringent multi-level review procedures must be followed before results are published.  There is a  tradition of publishing all results that have sensitivity to the effect under test without relying on the observed significance to determine whether or not to submit a manuscript or whether that manuscript is accepted.  Experimenters do not hesitate to publish null results.  Many of the proposed solutions to improve replicability of results in other fields have been, in some way or another, part of the culture of experimental particle physics for decades.
  
Particle physicists have long been cautioned about historical failures of even the most stringent checks and balances, and every new student is given examples of how well-meaning researchers can come to wrong conclusions because they misled themselves and therefore others.  The professional integrity of the collaboration members, the desire to deliver results that are correct within uncertainties and that are worth the effort and expense, and concern over repeating the mistakes of the past justify the rigor.  Scientists who have worked very hard on the many steps needed to conduct a large experiment do not want a relatively minor mistake in an analysis to produce a wrong result.  While by no means do all results in experimental particle physics meet the most rigorous standards, the techniques used to make the vast majority of them the best that can be produced are held as examples of good practices in science.

\subsection*{Follow-Up Discussion}

After the publication of this article in the Harvard Data Science Review
(\url{https://hdsr.mitpress.mit.edu/pub/1lhu0zvn/release/3?readingCollection=c6cf45bb}), an
interesting discussion followed.
Please see {\url{https://hdsr.mitpress.mit.edu/pub/32yz0u49/release/1}} for a
thoughtful comment from Andrew Fowlie, and {\url{https://hdsr.mitpress.mit.edu/pub/57tywz64/release/1}} for the authors' response.  The section title ``{\it Using $p$ Values to Quantify Discovery Significance}'' has been changed to ``{\it Using the p Value as a Tool for Discovery.}'' in this arXiv version.

\subsection*{Disclosure Statement}

Work supported by the Fermi National Accelerator Laboratory, managed and operated by Fermi Research Alliance, LLC under Contract No. DE-AC02-07CH11359 with the U.S. Department of Energy. The U.S. Government retains and the publisher, by accepting the article for publication, acknowledges that the U.S. Government retains a non-exclusive, paid-up, irrevocable, world-wide license to publish or reproduce the published form of this manuscript, or allow others to do so, for U.S. Government purposes.

The authors have no conflicts of interest to declare.

\subsection*{Acknowledgments}

The authors would like to thank Kyle Cranmer, Andrew Fowlie, Katri Lassila-Perini, Lara Lloret Iglesias, Alex Tapper and Nick Wardle for helpful discussions.

\appendix
\section{Glossary of Particle Physics Terminology}
\label{appendix:glossary}

The following glossary is a non-exhaustive list of terms used in  experimental particle physics.  An accessible introduction to particle physics is available at \url{https://particleadventure.org}.  More technical information, including extensive reviews, tables, and plots is available at {\url{https://pdg.lbl.gov}}.

\vskip 0.5cm

{\bf ALICE}:  A Large Ion Collider Experiment at CERN LHC.  This general-purpose detector is optimized to measure the collision products from heavy-ion collisions. 

{\bf ATLAS}:  A Toroidal LHC Apparatus.  This general purpose detector, like CMS, is optimized to measure the collision products from proton-proton collisions at the LHC.

{\bf antiparticles}:  Particles with the same mass, spin, and other properties, but with opposite electrical charge.  Some other properties, such as lepton number or baryon number, are also opposite to their matter particle counterparts. The antiparticle of the proton is the antiproton.

{\bf background}:  Processes that create events that are not from the desired signal process that contaminate samples of events that are intended to be enriched in a particular signal.

{\bf barn}:  A unit of cross section, equivalent to $10^{-24}$~cm$^2$.  Abbreviated b. A barn is approximately the cross sectional area of a uranium nucleus. See the entries for ``femtobarn,'' ``picobarn,'' ``nanobarn,'' ``microbarn,'' ``millibarn''  and ``cross section'' in this glossary.

{\bf barrel}:  The side of a cylindrical detector, shaped like a barrel.  Usually the largest part of a detector.  See the entries for ``end cap'' and ``forward detector'' in this glossary.

{\bf baryon}:  A hadron composed of three quarks (or 4 quarks and an anti-quark), held together by the strong force.  Protons and neutrons are common baryons.  Pentaquarks are also baryons.

{\bf baryon number}:  A quantum number that takes the value $+1$ for baryons and $-1$ for antibaryons.  Baryon number is observed to be conserved in known particle interactions. It is responsible for forbidding the carbon nuclei in our bodies decaying to six neutrons.  The dominance of matter over antimatter in the universe requires baryon-number nonconservation.

{\bf B hadron}:  A hadron containing a b quark.  Compared with protons, these are relatively massive, weighing 5 GeV or more, and have lifetimes of order $10^{-12}$ seconds.

{\bf b-jet}:  A jet containing at least one B hadron.

{\bf boson}:  A particle with an intrinsic spin that is an integer multiple of Planck’s constant divided by $2\pi$.  Named after Satyendra Nath Bose.  See the entry for ``fermion'' in this glossary.   Bosons follow Bose-Einstein statistics.  Photons, pions, $Z^0$ bosons and Higgs bosons are examples of bosons.  

{\bf BSM}:  Beyond the standard model

{\bf bubble chamber}:  A pressurized vessel containing a superheated cryogenic liquid, such as hydrogen, deuterium, neon, or other liquid.  Particles travel through the liquid, ionizing its molecules, nucleating bubbles.  These bubbles grow and are illuminated with a flash lamp and photographed.  A piston compresses the fluid, squeezing the gas in the bubbles back into the liquid stage, and then the pressure is lowered for the next exposure. Although they were  popular, they are now used very little.

{\bf calorimeter}:  A particle detector made out of a dense material in which high-energy particles interact, creating showers of newly-produced particles. See the entries for “electromagnetic calorimeter” and “hadron calorimeter” in this glossary.  The particles in the showers are detected by ionization or scintillation counters, and the total signal is proportional to the incident particle’s energy.

{\bf center-of-mass energy}:  The total energy of a system of particles in the frame of reference in which the total momentum is zero.  In a symmetric collider such as the LHC, this frame of reference is the laboratory frame in which the detectors are stationary.

{\bf CERN}:  Organisation Europ\'eenne pour la Recherche Nucl\'eaire.    The international high-energy physics laboratory in Geneva, Switzerland. 

{\bf CDF}:  The Collider Detector Facility at Fermilab, operated at the Tevatron proton-antiproton collider.  See the entry for D0 in this glossary.

{\bf chargino}:  One of several hypothetical particles, a quantum-mechanical mixture of the  supersymmetric partners of the $W^\pm$ bosons and the charged Higgs boson $H^\pm$.  A fermion with spin 1/2.  Symbols: $\widetilde \chi_1^\pm$ and $\widetilde \chi_2^\pm$, indicating the lighter and the heavier of the two charginos.

{\bf CMS}:  The Compact Muon Solenoid.  This general purpose detector, like ATLAS, is optimized to measure the collision products from proton-proton collisions at the LHC.

{\bf cosmic ray}:  A high-energy particle with extraterrestrial origin, impinging upon the atmosphere.  Often refers to the particles in the cascade resulting from its interaction with molecules in the atmosphere.  Cosmic rays consist of protons and heavier nuclei, and muons are the most common component of the cascades that reach the Earth’s surface.

{\bf cross section}:  A unit of area, characteristic of a specific interaction between particles, used to specify the interaction probability.  Usually symbolized as $\sigma$, and reported in b, mb, $\mu$b, nb, pb, or fb.  If a beam with $n$ particles per unit area impinges on a single-particle target with a cross section $\sigma$, then $n\sigma$ interactions are expected.  See the entries for ``femtobarn,'' ``picobarn,'' ``nanobarn,'' ``microbarn,'' ``millibarn,''  ``barn''  and ``luminosity'' in this glossary.

{\bf DESY}:  The Deutsches Elektronen-Synchrotron Laboratory in Hamburg, Germany.

{\bf D0}:  A collider detector at Fermilab, operated at the Tevatron proton-antiproton collider.  See the entry for CDF in this glossary.

{\bf dark energy}:  A hypothetical energy density in the vacuum, used to explain the accelerating rate of expansion of the universe.

{\bf dark matter}:  A hypothetical substance that interacts gravitationally with particles of the standard model.  Dark matter may interact also via the weak interaction, but it lacks electromagnetic and strong interactions.   Dark matter is used to explain galactic rotation curves, large-scale structure in the universe, the observed strength of gravitational lensing, and properties of the cosmic microwave background radiation.  Dark-matter particles are actively sought in laboratories around the world but at the time of writing, there is no direct evidence for them.

{\bf efficiency}:  The probability that an event selection algorithm will correctly select an event of the desired class.

{\bf electromagnetic calorimeter}:  A calorimeter intended to measure $e^+e^-$ cascades initiated by high-energy photons or electrons incident on a dense material.

{\bf electron volt}:  The amount of energy given to one electron as it travels across a potential difference of one volt.

{\bf end cap}:  Detectors that fill the angular coverage between the barrel and the forward detectors.  End-cap detectors cover the open ends of the barrel.  End caps typically can be opened when the accelerator is not operating so that detector components in the barrel and the end caps be serviced.

{\bf event}:  A collision of two particles and all subsequent interactions and decays of particles produced in the collision.  

{\bf femtobarn}:  $10^{-15}$ barns.  See the entries for ``picobarn,'' ``nanobarn,'' ``microbarn,'' ``millibarn,''  ``barn''  and ``cross section'' in this glossary.

{\bf fermion}:  A particle with half-odd integer spin.  Electrons, protons, and neutrons are examples of fermions.  Identical fermions obey the Pauli exclusion principle -- two fermions cannot be in the same quantum state at the same time.  Fermions obey Fermi-Dirac statistics.  Named after Enrico Fermi.

{\bf flavor}:  A name given to a description of what type of quark or lepton a particle is.  There are six quark flavors: up, down, strange, charm, top and bottom. There are three  charged lepton flavors:  $e$, $\mu$, and $\tau$.  There are also three neutrino flavors:  $\nu_e$, $\nu_\mu$, and $\nu_\tau$.

{\bf forward detector}  Detector components that measure particles that travel at very small angles with respect to the beam axis in a collider detector.  Forward detectors may contain tracking detectors, calorimeters, muon detectors, or any combination of these.  See the entries for ``barrel'' and ``end cap'' in this glossary.

{\bf GeV}:  One billion electron volts.  While this is a unit of energy, physicists often refer to masses in GeV via $E=mc^2$, in units where $c=1$.

{\bf gluon}:  The boson that transmits the strong nuclear force.  It interacts with quarks, gluons, and particles composed of them.  Symbol: $g$.

{\bf gluino}  A hypothetical particle, the supersymmetric partner of the gluon.  A fermion with spin 1/2.  Symbol: $\widetilde g$.

{\bf graviton}:  A hypothetical quantum excitation of the gravitational field.  Gravitons have not yet been observed.  They are predicted to be massless, spin-2 bosons.

{\bf gravitino}: A hypothetical particle, the supersymmetric partner of the graviton.  A fermion with spin 3/2.  Symbol: $\widetilde G$.

{\bf hadron}:  a composite particle made up of quarks, and held together with the strong force.

{\bf hadron calorimeter}:  A detector made out of a dense material, typically located farther from a collider’s interaction point than the electromagnetic calorimeter.  Electromagnetic showers are largely contained in the electromagnetic calorimeter, while hadrons can punch through the electromagnetic calorimeter and interact in the hadron calorimeter.

{\bf Heisenberg uncertainty principle}:  A property of quantum mechanics that precludes a quantum state from simultaneously having well-defined values for two conjugate variables.  Two examples of conjugate variable pairs are position and momentum, and energy and time.  Quantitatively, the products in the spreads in the distributions of the variables must exceed $h/4\pi$: $\Delta x\Delta p \ge h/4\pi$ and $\Delta E\Delta t>h/4\pi$, where $x$ is a particle's location, $p$ is its momentum, $E$ is its energy, $t$ is the time over which the energy is measured, and $h$ is Planck's constant.  See the entry for ``virtual particle'' in this glossary.

{\bf Higgs boson}:  A quantized excitation of the Higgs field.  A boson with spin zero and a mass of approximately 125 GeV.

{\bf Higgs field}:  A field with a nonzero average value in the vacuum.  Responsible for the nonzero masses of elementary particles, via the Higgs mechanism.

{\bf Higgs mechanism}:  An explanation by which the particles in the standard model acquire mass from spontaneous symmetry breaking in the Higgs field.  Predicted in 1964 in order to reconcile the nonzero masses of particles with the symmetries of the standard model, its existence was confirmed with the discovery of the Higgs boson in 2012.

{\bf ionization}:  The ejection of atomic electrons, often due to the passage of high-energy charged particles.  Ionization is useful in the detection of charged particles as the liberated electrons, and sometimes the remaining ions, can be detected electronically.

{\bf integrated luminosity}:  A measure of the size of a data set in a collider experiment.  Expressed in terms of inverse cross section.  A dataset with a size of 35.9~fb$^{-1}$ is expected to produce 35.9 events of a type if the cross section of events of that type is 1~fb.  See the entries for ``cross section,'' ``luminosity,'' and ``femtobarn'' in this glossary. 

{\bf jet}:  A cluster of particles emitted in a narrow cone during a collision of high-energy particles.

{\bf kaons}: Charged or neutral mesons containing a strange quark and either an up antiquark or a down antiquark, respectively.  Symbols:  $K^-$ and $K^0$, with antiparticles $K^+$ and $\overline K^0$.

{\bf keV}:  kilo electron volt.

{\bf LAr}:   Liquid argon.  Used in particle detectors because argon atoms ionize when charged particles pass through, and the drifting electrons can be collected electronically.

{\bf LEP}:  The Large Electron-Positron Collider at CERN.  Beams of counter-rotating electrons and positrons of energy up to 104.5 GeV collided at four interaction points, corresponding to center-of-mass energies of up to 209~GeV.  The four detectors on the LEP ring were ALEPH, DELPHI, L3 and OPAL. This accelerator predated the LHC, which was situated in the LEP tunnel after LEP was dismantled

{\bf lepton}:    Leptons are currently believed to be fundamental particles.  They have spin 1/2 and are thus fermions.  The charged leptons are the electron, the muon, and the tau.  The neutral leptons are the three neutrinos, $\nu_e$, $\nu_\mu$ and $\nu_\tau$.

{\bf lepton number}:  A quantum number that takes the value $+1$ for leptons like electrons, muons, or their neutrinos, and $-1$ for their antiparticles.  See the entry for ``baryon number'' in this glossary.  Lepton number is also observed to be conserved in known particle interactions.

{\bf LHC}:  The Large Hadron Collider at CERN.  It is a ring of superconducting magnets 27 km in circumference.  It accelerates counter-rotating proton beams with energies of 6.5 TeV for a total center-of-mass energy of 13 TeV. It has also collided lead nuclei.

{\bf LHCb}:  A general-purpose detector at the LHC, optimized for measuring the decay products of B hadrons.

{\bf LSND}:  The Liquid Scintillator Neutrino Detector at the Los Alamos National Laboratory.

{\bf LSP} Lightest supersymmetric particle.  Often assumed to be stable, a neutral LSP, such as a neutralino, is a candidate component of  dark matter.

{\bf luminosity}:  The instantaneous rate of production of collision data in a collider.  Luminosity is usually expressed in cm$^{-2}/s$.  The product of luminosity and cross section is the event production rate.  See the entries for ``femtobarn'' and ``cross section'' in this glossary.

{\bf meson}:  A hadron consisting of a quark and an antiquark, bound together with the strong nuclear force. They have integer spin, and hence are bosons. Pions and kaons are examples of mesons.

{\bf MeV}:   $10^6$ electron volts.   While this is a unit of energy, physicists often refer to masses in MeV via $E=mc^2$, in units where $c=1$. The masses of the electron and the proton are about half an MeV and 938 MeV, respectively.

{\bf microbarn}:  $10^{-6}$ barns.  Abbreviated $\mu$b.  See the entries for ``femtobarn,'' ``picobarn,'' ``nanobarn,'' ``millibarn,''  ``barn''  and ``cross section'' in this glossary.

{\bf millibarn}:  $10^{-3}$ barns.  Abbreviated mb.  See the entries for ``femtobarn,'' ``picobarn,'' ``nanobarn,'' ``microbarn,''  ``barn''  and ``cross section'' in this glossary.

{\bf MiniBooNE}:  An experiment at Fermilab designed to observe short-baseline neutrino oscillations.  BooNE stands for `Booster Neutrino Experiment.'

{\bf ML}:  Machine Learning.  MVA techniques need to provide optimal mappings of the input variable values to useful classification, discriminant or regression variables.  Discriminant, classification, and regression functions are optimized by training them on sets of labeled examples of signal events and background events.  The training procedures optimize the performance of the MVAs on the training samples.  The resulting trained MVA functions are then used to interpret unlabeled data.

{\bf muon}:  A charged lepton, with properties similar to that of an electron, but with a mass of 106 MeV.  Symbol: $\mu^-$.  High-energy muons create trails of ionized molecules in detectors but they have a much smaller probability of initiating showers of secondary particles compared with electrons, and thus they penetrate thick layers of material in particle detectors, such as the electromagnetic and hadronic calorimeters.

{\bf muon detector}:  A component of a collider detector located outside the hadron calorimeter.  Usually made of steel interleaved with ionization or scintillation detectors, nearly all particles that are not stopped by the calorimeters are muons.

{\bf MVA}:  Multivariate Analysis.  Often many quantities can be measured for a single event, and a decision must be made based on the values of these quantities, such as whether or not to select an event for inclusion in an analysis, or a quantity must be estimated from the set of measured values.  MVA techniques distill the information from multiple variables into a single variable or a classification choice.  See the entry for ``ML'' in this glossary.

{\bf nanobarn}:  $10^{-9}$ barns.  Abbreviated nb. See the entries for ``femtobarn,'' ``picobarn,'' ``microbarn,'' ``millibarn,''  ``barn''  and ``cross section'' in this glossary.

{\bf neutralino}:  One of several hypothetical particles, a quantum-mechanical mixture of the  supersymmetric partners of the photon, the $Z^0$ boson, and the Higgs bosons, of which there are two neutral ones in the minimal supersymmetric model.  A fermion with spin 1/2.  Symbols: $\widetilde \chi_1^0$ ... $\widetilde \chi_4^0$, indicating the four mass states.

{\bf neutrino oscillations}:  A phenomenon in which neutrinos of one flavor spontaneously transform into neutrinos of other flavors and back again.

{\bf NLO}:  Next-to-leading order. A theoretical calculation of a cross section typically includes the process of interest plus radiative corrections, to account for gluon exchange and gluon emission, for example.  A NLO calculation is a truncated sum containing the first term in the Taylor series describing radiative corrections.

{\bf NLSP}  Next-to-lightest supersymmetric particle.  In many models, the LSP is assumed to be stable and the end of a decay chain of heavier supersymmetric particles.  The NLSP may decay to the LSP and visible SM particles, providing a signature for search analyses to use.

{\bf off-resonance running}  Typically used in $e^+e^-$ colliders, this is a running mode in which the beam energy is tuned so that a particular particle, such as a $Z^0$ boson, is produced much less often than when the beam energy is tuned so the center-of-mass energy matches the particle's rest mass.

{\bf pentaquark}:  A hadron made up of four quarks and an antiquark.  Pentaquarks are fermions with baryon number +1.

{\bf PEP}:  A circular electron-positron collider that was at SLAC.  Five detectors were installed:  HRS, TPC-2$\gamma$, Mark-II, MAC and DELCO.

{\bf PETRA}:  A circular electron-positron collider that was at DESY.  Four detectors were installed:  Mark J, TASSO, PLUTO and JADE.

{\bf photon}:  A quantum excitation of the electromagnetic field.  Transmits the electromagnetic force.  Photons are bosons with spin 1. Symbol: $\gamma$.

{\bf picobarn}:  $10^{-12}$ barns.  Abbreviated pb. See the entries for ``femtobarn,'' ``nanobarn,'' ``microbarn,'' ``millibarn,''  ``barn''  and ``cross section'' in this glossary.

{\bf pion}:  The lightest meson.  May be charged or neutral. Symbols: $\pi^+$, $\pi^-$, $\pi^0$. Most tracks in hadron collider events are caused by charged pions.  Pions are exchanged between nucleons in the nucleus, holding them together.  Pion exchange was originally called the strong nuclear force, but it later was realized to be an emergent phenomenon, arising from the underlying quark and gluon interactions.

{\bf pixel detector}:  A two-dimensional particle detector segmented into small sensitive areas.  Pixel detectors measure three-dimensional track trajectories in one of two ways.  Multiple layers of pixel detectors which detect ionization within the pixels can be used to measure points along a track.  Alternatively, ionization electrons may drift through a gaseous or liquid medium to a single plane of pixels; the drift time provides the measurement of the third dimension.

{\bf positron}:  An anti-electron.  It has spin 1/2 and charge $+e$, while an electron has charge $-e$.  Symbol:  $e^+$.

{\bf purity}:  The fraction of a selected sample of events that originate from a desired signal process.  The remainder come from background sources.  This classification assumes a mixture model for events where rates of signals and backgrounds add arithmetically.  In some cases, the signal and the background processes interfere quantum mechanically and purity is not well defined.

{\bf quark}:  An elementary constituent of hadrons.  Quarks are fermions with spin 1/2.  They interact with gluons and are bound up in pairs, or groups of three or more.  Up-type quarks have charge $+2e/3$ and down-type quarks have charge $-e/3$.

{\bf scintillation}:  A charged particle may excite a molecule to a higher-energy state.  The decay of this higher-energy state often is accompanied by the emission of one or more scintillation photons.  Scintillation is a common phenomenon used in particle detectors as scintillation photons are easily detected with photomultiplier tubes or more modern solid-state equivalents.

{\bf selectron}:  A hypothetical spin-zero supersymmetric partner of the electron. Symbol: $\widetilde e^-$. Anti-selectrons correspond to positrons.

{\bf signal}:  A source of events that are of interest and are the subject of study in a particular analysis.  See the entry for ``background'' in this glossary.

{\bf SLAC}:  The Stanford Linear Accelerator Center in Menlo Park, California, USA.  Now called SLAC NAL, for SLAC National Accelerator Laboratory.

{\bf slepton}  A class of hypothetical particles, the supersymmetric partners of the leptons.  A boson with spin 0.  Examples are selectrons, smuons, staus and sneutrinos.

{\bf solenoid magnet}:  A cylindrical electromagnet.  In a collider detector, the coils are usually superconducting.  The magnetic field is provided so that the Lorentz forces on the moving charged particles causes their trajectories to bend, and hence enables the particle’s momentum to be determined.

{\bf sneutrino}  The three sneutrinos are the supersymmetric partners of the neutrinos.  They are electrically neutral bosons with spin 0.

{\bf squark}  The six squarks are the supersymmetric partners of the quarks.  They are bosons with spin 0 and they have the same electrical charges as the corresponding quarks.

{\bf standard model}:  A description of the fundamental particles and interactions of particle physics, except for gravity. It has been very successful in interpreting high energy physics data.  Abbreviated SM.

{\bf sterile}:   A term used to describe a neutrino that does not interact via the weak nuclear force.

{\bf superconductor}:  A material that exhibits zero electrical resistance when cooled below its superconducting transition temperature.  Magnetic fields are expelled from the bulk of superconductors, differentiating them from merely good conductors.

{\bf supersymmetry}:  A hypothetical symmetry between fermions and bosons.  Each fermion of the standard model has a bosonic partner with similar properties if this hypothesis is true, and each boson has a fermionic partner.  While supersymmetry has many attractive properties, such as stabilizing the mass of the Higgs boson against radiative corrections and providing a dark-matter candidate,  no evidence has yet been observed supporting it.

{\bf TeV}:  A tera electron volt.  $10^{12}$ electron volts.

{\bf Tevatron}:  A former proton-antiproton collider at Fermilab.  Each beam had an energy of 980 GeV, for a center-of-mass energy of 1.96 TeV.

{\bf tile calorimeter}:  The hadron calorimeter of ATLAS, consisting of a sandwich of steel plates and scintillating tiles.

{\bf time projection chamber}:  A tracking detector in which a liquid or gas medium resides in a strong electric field.  Electrons liberated from molecules ionized by high-energy particles drift towards wire- or pixel-based detectors on one end.  The time taken by electrons to drift through the medium provides a measure of the distance between the detection element and the location at which the ionization occurred.

{\bf toroid magnet}:  A magnet in which the field lines are guided in loops around an axis.  ATLAS’s muon system contains large toroidal magnets.

{\bf TPC}:  Time projection chamber.  See the entries for ``tracking detector'' and ``pixel detector'' in this glossary.

{\bf track}:  The trail of ionized atoms or molecules left by the passage of a high-energy charged particle in a tracking detector.

{\bf tracking detector}:  A detector designed to measure the trajectories of charged particles via their ionization signatures.  Examples of tracking detectors include bubble chambers, wire drift chambers, silicon strip and pixel vertex detectors, and pixel time projection chambers.

{\bf transition radiation tracker}:  A tracking detector that takes advantage of the electromagnetic shock wave incurred when a high-energy particle leaves a transparent medium with one index of refraction and enters a medium with another.

{\bf trigger}:  Logic designed to decide when to read out the data from the detector, for permanent storage. Triggers are designed to search for “interesting” events and not to fill up data-storage resources with uninteresting data.

{\bf vertex detector}:  A tracking detector with very fine granularity located within a few tens of cm of a collider’s interaction region, in the center of a collider detector.  Commonly made out of silicon strips or pixels, the position resolution is typically of order of a few microns.

{\bf virtual particle}  A particle with temporary existence whose effects can be inferred by its interactions.  The energy required to create a force-carrying particle such as a $W^\pm$ boson may be greater than the energy available in a particular decay, but the $W^\pm$ boson will still cause the interaction to occur.  The energy imbalance is allowed by the Heisenberg uncertainty principle if it is short-lived enough.  Most forces experienced by subatomic particles are transmitted by virtual particles.

\begin{figure}[ht]
\begin{center}
\includegraphics[width=0.3\textwidth]{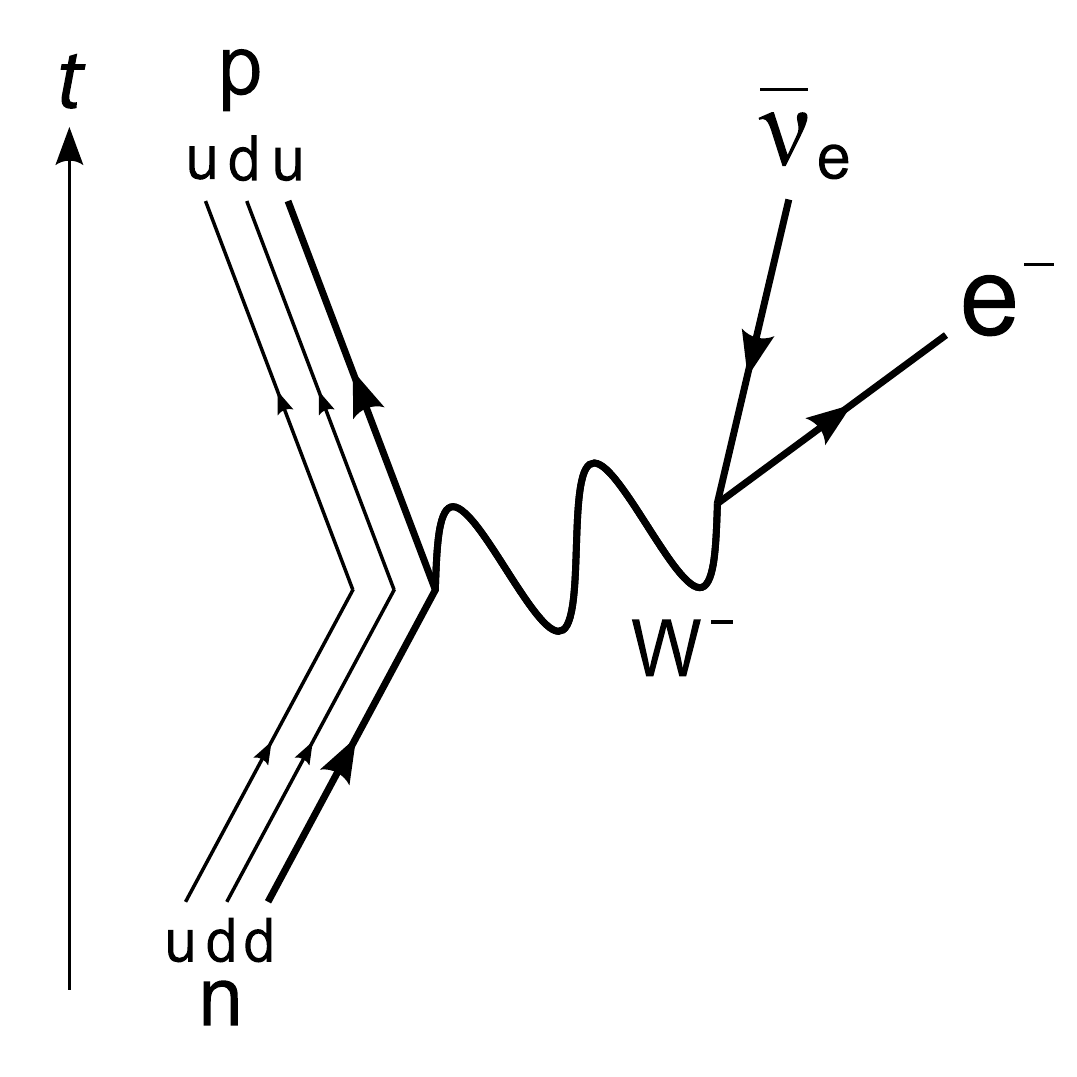} \\
\end{center}
\caption{Diagram showing neutron beta decay via $W^-$ boson exchange.  Time advances along the vertical axis.  Image credit:  Joel Holdsworth, Wikipedia.}
\label{fig:neutronbetadecay}
\end{figure}

\begin{boldmath}{\bf $W^\pm$ boson}\end{boldmath}:  One of the particles transmitting the weak nuclear force.  $W^\pm$ bosons have spin 1 and charge $\pm 1$.  A particle interacting with a $W^\pm$ boson must change its electric charge by one unit, thereby changing it into another type of particle.  Figure~\ref{fig:neutronbetadecay} shows an example of an interaction mediated by a $W^-$ boson, neutron beta decay.  A neutron emits a virtual $W^-$ boson, thereby turning into a proton.  The $W^-$ boson then produces an electron and an electron anti-neutrino.   $W^\pm$ bosons have a mass of approximately 80.3 GeV.

\begin{boldmath}{\bf $Z^0$ boson}\end{boldmath}:  The neutral transmitter of the weak nuclear force. $Z^0$ bosons have spin 1 and a mass of approximately 91.2 GeV.


\bibliographystyle{apalike}
\bibliography{references}

\end{document}